\documentclass[letterpaper,twocolumn,10pt]{article}
\usepackage{usenix}
\usepackage{graphicx}
\usepackage{amsmath}
\usepackage{hyperref}
\usepackage{booktabs}
\usepackage{tikz}
\usepackage{enumitem}
\usepackage{float}
\usepackage{authblk}
\usepackage{tabularx}
\usepackage{xcolor}
\usepackage{xspace}
\usepackage{multirow}
\usepackage{subcaption}
\usepackage{filecontents}
\usepackage[normalem]{ulem}
\usepackage{ifthen}
\usepackage{amssymb}   
\usepackage{algorithm}
\usepackage{algpseudocode}

\PassOptionsToPackage{usenames,dvipsnames}{color} 
\hypersetup{unicode=true,
            colorlinks=true,
            linkcolor=blue,
            citecolor=blue,
            anchorcolor=blue,
            urlcolor=blue,
            breaklinks=true}
\setlist{topsep=1pt, partopsep=0pt, itemsep=1pt, parsep=1pt, leftmargin=19pt}

\usepackage{titlesec}
\titlespacing*{\subsection}{0pt}{4pt plus 3pt minus 2pt}{1pt plus 3pt minus 1pt}
\titlespacing*{\subsubsection}{0pt}{4pt plus 3pt minus 2pt}{0pt plus 3pt minus 1pt}

\makeatletter
\def\maxwidth{\ifdim\Gin@nat@width>\linewidth\linewidth\else\Gin@nat@width\fi}
\def\maxheight{\ifdim\Gin@nat@height>\textheight\textheight\else\Gin@nat@height\fi}
\makeatother
\setkeys{Gin}{width=\maxwidth,height=\maxheight,keepaspectratio}
\makeatletter
\g@addto@macro{\UrlBreaks}{\UrlOrds}
\makeatother

\newcommand\paraspace{\vspace*{0.5ex}}
\providecommand\parab[1]{\paraspace\noindent\textbf{#1}\xspace\xspace}

\newtoggle{reviewmode}
\newtoggle{anony}

\newcommand{\projurl}[1]{%
    \iftoggle{anony}{%
        URL is hidden for review purpose%
    }{%
        \url{#1}%
    }%
}

\definecolor{teal}{rgb}{0.0, 0.5, 0.5}
\definecolor{olive}{rgb}{0.5, 0.5, 0.0}
\definecolor{pink}{rgb}{1.0, 0.75, 0.8}
\definecolor{lightgray}{gray}{0.75}
\definecolor{mediumgray}{gray}{0.5}
\definecolor{darkgray}{gray}{0.25}
\definecolor{charcoal}{gray}{0.2}
\definecolor{turquoise}{rgb}{0.25, 0.88, 0.82}
\definecolor{coral}{rgb}{1.0, 0.5, 0.31}
\definecolor{navyblue}{rgb}{0.0, 0.0, 0.5}
\definecolor{lime}{rgb}{0.75, 1.0, 0.0}
\definecolor{darkgreen}{rgb}{0.0, 0.5, 0.0}
\definecolor{violet}{rgb}{0.56, 0.0, 1.0}
\definecolor{lightgreen}{rgb}{0.85, 1.0, 0.85}

\definecolor{burgundy}{cmyk}{0.5, 1.0, 0.7, 0.4}
\definecolor{olivegreen}{cmyk}{0.64, 0, 0.95, 0.4}
\definecolor{peach}{cmyk}{0, 0.5, 0.7, 0}
\definecolor{mustard}{cmyk}{0, 0.3, 1, 0}

\newcolumntype{N}{>{\raggedleft\arraybackslash}p{1.4cm}} 
\newcolumntype{U}{>{\raggedright\arraybackslash}p{0.8cm}} 

\toggletrue{anony}  

\newcommand{\yanqi}[1]{\textcolor{blue}{Yanqi: #1}}

\newcommand{\name}{AMoE\xspace}

\newtoggle{showmarks}
\toggletrue{showmarks}
\iftoggle{showmarks}{
\newcommand{\seojin}[1]{{\footnotesize[\textcolor{orange}{\sf\textit{#1 - Seojin}}]}}

\newcommand\red[1]{\textcolor{red}{#1}}
\newcommand\redstrike[1]{\red{\sout{#1}}}
\newcommand\green[1]{\textcolor{\green}{#1}}
\newcommand\greenstrike[1]{\green{\sout{#1}}}
\newcommand\orange[1]{\textcolor{orange}{#1}}
\newcommand\orangestrike[1]{\orange{\sout{#1}}}
\newcommand\blue[1]{\textcolor{blue}{#1}}
\newcommand\bluestrike[1]{\blue{\sout{#1}}}
\newcommand\purple[1]{\textcolor{purple}{#1}}
\newcommand\purplestrike[1]{\purple{\sout{#1}}}
}{
\newcommand{\seojin}[1]{}
\newcommand{\gwkim}[1]{}
\newcommand{\yanqi}[1]{}
\newcommand\red[1]{#1}
\newcommand\redstrike[1]{\unskip}
\newcommand\green[1]{#1}
\newcommand\greenstrike[1]{\unskip}
\newcommand\orange[1]{#1}
\newcommand\orangestrike[1]{\unskip}
\newcommand\blue[1]{#1}
\newcommand\bluestrike[1]{\unskip}
\newcommand\purple[1]{\unskip}
\newcommand\purplestrike[1]{\unskip}
}

\begin{document}

\date{}
\title{Toward Cost-Efficient Serving of Mixture-of-Experts with Asynchrony}

\newboolean{anonymizeauthors}
\setboolean{anonymizeauthors}{false}  

\ifthenelse{\boolean{anonymizeauthors}}{
    \author{Paper\# 1058}
}{
    \author{
    Shaoyu Wang$^{1}$\textsuperscript{*} \quad
    Guangrong He$^{1}$\textsuperscript{*} \quad
    Geon-Woo Kim$^{2}$ \quad
    Yanqi Zhou$^{3}$ \quad
    Seo Jin Park$^{1}$ \\
    $^{1}$University of Southern California \quad
    $^{2}$University of Texas at Austin \quad
    $^{3}$Google
    }
}

\maketitle

\ifthenelse{\boolean{anonymizeauthors}}{
} {
\renewcommand{\thefootnote}{}
\footnotetext{\textsuperscript{*}Equal contribution}
\renewcommand{\thefootnote}{\arabic{footnote}}
}

\pagestyle{plain}
\begin{abstract}

Mixture-of-Experts (MoE) architectures offer the promise of larger model capacity without the prohibitive costs of fully dense designs. However, in real-world inference serving, load skew across experts often leads to suboptimal device utilization and excessive synchronization overheads. This paper introduces Asynchronous Expert Parallelism (AEP), a new paradigm that decouples layer execution from barrier-style synchronization. By dynamically queuing tokens at each layer (referred to as $\mu$-queuing) and adaptively re-batching them on demand, GPUs avoid waiting for straggling experts and instead continuously process whichever layer is ready. This asynchronous approach mitigates two major inefficiencies in traditional expert-parallel systems: (1) idle GPU time while waiting for the hottest expert, and (2) small-batch executions on colder experts that waste memory bandwidth. 

We implement these ideas in a serving system called \name, which disaggregates attention from expert layers and uses a defragging scheduler to reduce batch fragmentation. Evaluations on prototype MoE models show that \name improves throughput by up to 2.7x compared to state-of-the-art baselines, incurring a manageable latency penalty and providing a cost-effective operating point. Furthermore, experiments demonstrate nearly linear scalability to multi-node settings, whereas the baseline system shows no throughput increase even when the number of GPUs is doubled. 

\end{abstract}

\section{Introduction}

It is well known that the accuracy of a DNN (including LLM) is dependent on the model size~\cite{palm}, so high-performance models~\cite{gpt4, gemini} are rumored to use more than 1.5 trillion parameters. Unfortunately, such scaling of models increases the serving costs. For example, open AI charges \$150 for 1 million token generation with GPT-4.5~\cite{openai-pricing}, prohibitively expensive for everyday applications. 

To enable scaling of model sizes without increasing the amount of computation for serving, Mixture-of-Experts (MoE) models are receiving increasing attention~\cite{outrageously, switch, glam, mixtral, megablocks, grok, deepseek, dbrx}. MoE models are composed of many specialized experts, only a few of which (e.g., 1-2) are activated for each token, greatly reducing the amount of computation required for each token. In theory, we can increase the model size for better accuracy without increasing the amount of computation, and it's proven to reduce training cost greatly~\cite{switch}.

\begin{figure}
\centering
\includegraphics[width=0.85\columnwidth]{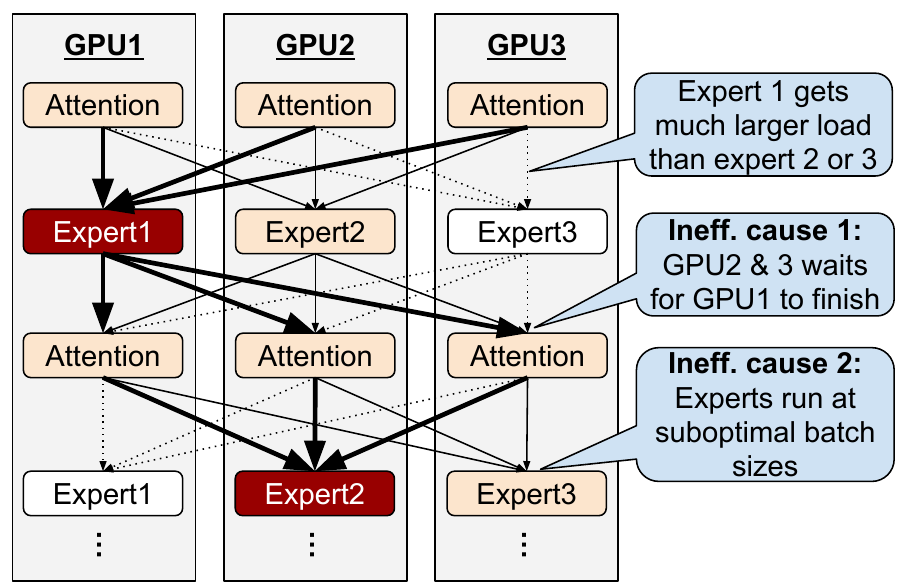}
\caption{\label{fig:problem} Expert load skews causes inefficient GPU executions.}
\end{figure}


However, unlike training, today's cost of MoE serving is still suboptimal because of the load skews across experts~\footnote{During training, expert loads are self-balanced with a loss function tweak~\cite{outrageously}. However, the workloads during serving are different from training, resulting in significant load skews across experts~\cite{lina}}.
As shown in Fig.~\ref{fig:problem}, the load skew causes two serving efficiency challenges: (1) accelerator stalling when experts are sharded across GPUs~\cite{lina} and (2) sub-optimal batch sizes for expert layer computations. Many MoE systems, such as SwitchTransformer~\cite{switch}, DeepSpeed-MoE~\cite{deepspeed}, DeepSeek~\cite{deepseek} and GLaM~\cite{glam}, shard experts across GPUs to fit large MoE models (expert parallelism). In such sharded deployment, GPUs in charge of cold experts will get lower loads and will be stalling while waiting for the slowest expert to finish. 
In addition to GPU stalls, expert load skew also hurts GPU efficiency by preventing layers' executions at optimal batch sizes; cold expert computations are heavily bottlenecked by the GPU's High Bandwidth Memory (HBM) bandwidth for loading parameters, while hot experts run at too large of a batch which hurts latency without any throughput benefits.

These inefficiencies arise because today's serving systems batch multiple requests and execute the fixed batch through all layers. 
With the rigid batching across all layers, all-to-all barrier-style communication before and after expert layers is inevitable and causes inefficiency when loads are not perfectly balanced.
Strawman approaches like matching the skewed loads by provisioning more GPUs for hot experts won't work well enough since expert load skews are known to shift dynamically~\cite{lina, huang2024toward, he2025capacityaware, cong2024prediction}.

\begin{figure}
\centering
\includegraphics[width=\columnwidth]{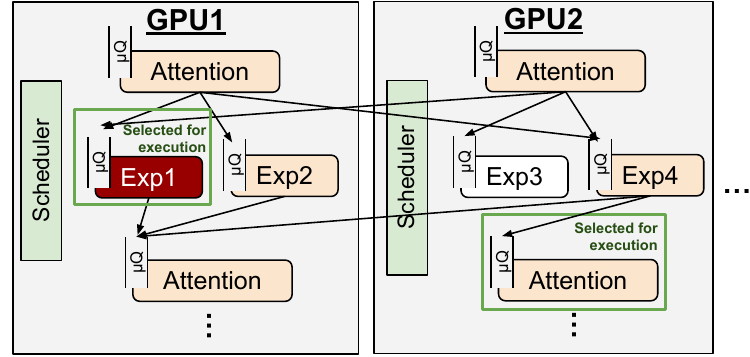}
\caption{\label{fig:aep} Asynchronous Expert Parallelism. Schedulers in each GPU freely selects layer to execute with accumulated tokens.}
\end{figure}

We propose to solve the efficiency challenges in MoE serving via \textbf{Asynchronous Expert Parallelism (AEP)}, where each device can execute and forward output independently in an asynchronous manner (Figure~\ref{fig:aep}). The key technique is layer-wise scheduling: queuing tokens at the granularity of individual layers~(which we call $\mu$-queuing) and adaptively re-batching and executing just in time with the tokens so far accumulated at the layer's own $\mu$-queue. Due to adaptive re-batching, GPUs do not need to wait for barrier-style all-to-all communication to finish. Instead, they stay busy as long as enough load is offered at any layer. 
By colocating more than one expert layer on a GPU, scheduler can multiplex layers to prioritize execution of hot experts with enough input tokens and let cold experts to accumulate more tokens before execution. 

To demonstrate the efficacy of AEP, we built a prototype MoE serving, \name. With a small scale (8 experts, 8 GPUs) expert-compute-heavy workloads, our approach improved throughput up to 2.7x from the state of the art serving system with expert parallelism support (SGLang\cite{sglang}), with a penalty on higher inter-token latency. On an extended scale (16 experts, 16 GPUs), AEP showed almost linear scaling of throughput while SGLang with standard EP showed no throughput increase when scaled from 8 GPU settings.

We make following contributions:
\begin{itemize}
    \item We propose a new parallel serving method, asynchronous expert parallelism, which can avoid many of limitations of expert parallelism while retaining the its benefits of better scalability with low communication overheads.
    \item We design and implement a new MoE serving system, \name, which supports asynchronous expert parallelism (AEP). We address several challenges in realizing AEP: (1) token-level dependency tracking (2) queuing delay minimization (3) high performance asynchronous communication.
    \item We characterize workloads that can most benefit from AEP.
    \item We open-source our serving system, \name, for public use.
\end{itemize}

\section{Background and Motivation}        


\subsection{Efficiency Challenge in MoE Serving}

The Mixture-of-Experts (MoE) architecture enhances the efficiency and scalability of large language models by selectively activating only a subset of specialized sub-models, called experts, for each input. A routing layer determines which experts are most relevant, allowing for a significant reduction in computational overhead while enabling the training of models with billions of parameters. By integrating specialized expert layers into transformer architectures, MoE achieves higher performance without increasing computing costs than dense models.

Early MoE research focused on leveraging this sparsity for improved accuracy without increasing FLOPs per token~\cite{switch, stmoe, glam, expertchoice}.
For instance, the Switch Transformer~\cite{switch} explored scaling to a large number of experts (up to 2048) while maintaining the same activated parameter size, demonstrating that this could yield higher model accuracy within the same training compute budget. To effectively manage and utilize such numerous experts, expert parallelism (EP) was introduced, distributing individual experts across different hardware devices (e.g., GPUs). This distribution strategy is crucial for enabling computationally feasible training and inference with large-scale MoE models. 

However, a key practical challenge arises with expert parallelism: \textit{expert load imbalance}. Tokens within a processing batch are often routed unevenly, causing some experts (and their corresponding devices) to receive significantly more tokens than others. During training, this imbalance was often considered manageable. Mitigation techniques such as auxiliary load balancing losses~\cite{lepikhin2021gshard, shazeer2018}, strategies allowing experts to drop excess tokens~\cite{glam}, routing by expert's choice~\cite{expertchoice} and the use of very large batch sizes helped suppress the negative impacts of load skew.

Since the widespread deployment of large models beginning around 2023, the operational cost of serving inferences has gained critical importance, often outweighing training costs. In this serving context, expert load imbalance poses a serious impediment to efficiency when using EP. The load distribution patterns observed during serving often differ significantly from those during training and can shift dynamically based on the input workload characteristics, nullifying training-time optimizations like auxiliary losses~\cite{lina}. Typically, serving workloads exhibit both persistent load skew (some experts are consistently favored) and transient, request-dependent skew~\cite{lina, huang2024toward, he2025capacityaware, cong2024prediction}. 

\begin{figure}
    \centering
    \includegraphics[width=\columnwidth]{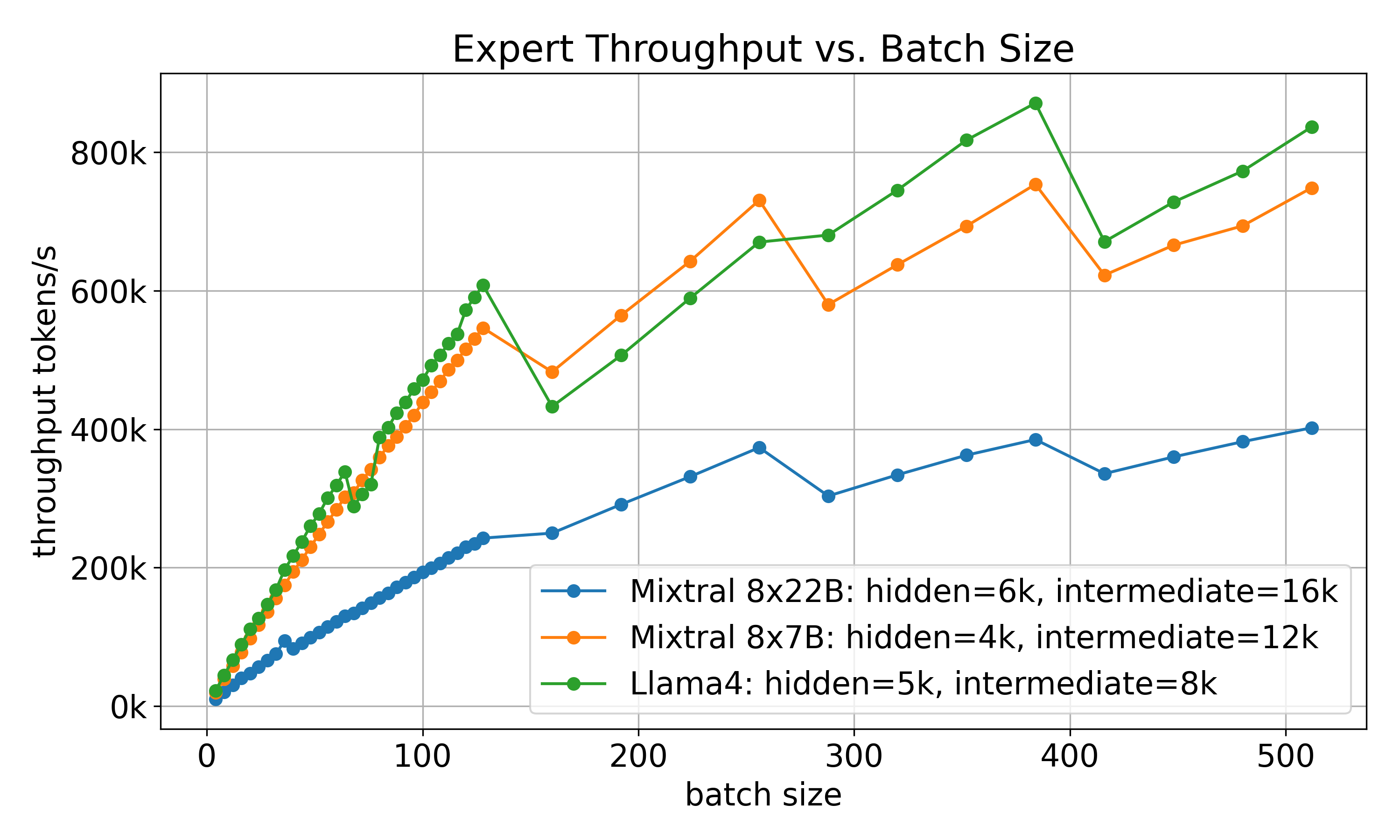}
    \caption{Execution throughput of a single expert layer with different batch sizes on A100 40GB.}
    \label{fig:tputByBatch}
\end{figure}

This expert load skew introduces two primary efficiency challenges during MoE serving with EP. First, it \textbf{prevents expert layer computations from running at optimal batch sizes on all devices.} Experts receiving few tokens ("cold experts") operate with small batch sizes. At these low batch sizes, GPU computation is often bottlenecked by the time taken to load expert weights from High Bandwidth Memory (HBM), leading to underutilization of the GPU's computational units. Figure~\ref{fig:tputByBatch} shows that increasing batch size increases throughput almost linearly until the batch size of 128, suggesting that any executions with smaller batches are wasteful.

\begin{figure}
    \centering
    \includegraphics[width=\columnwidth]{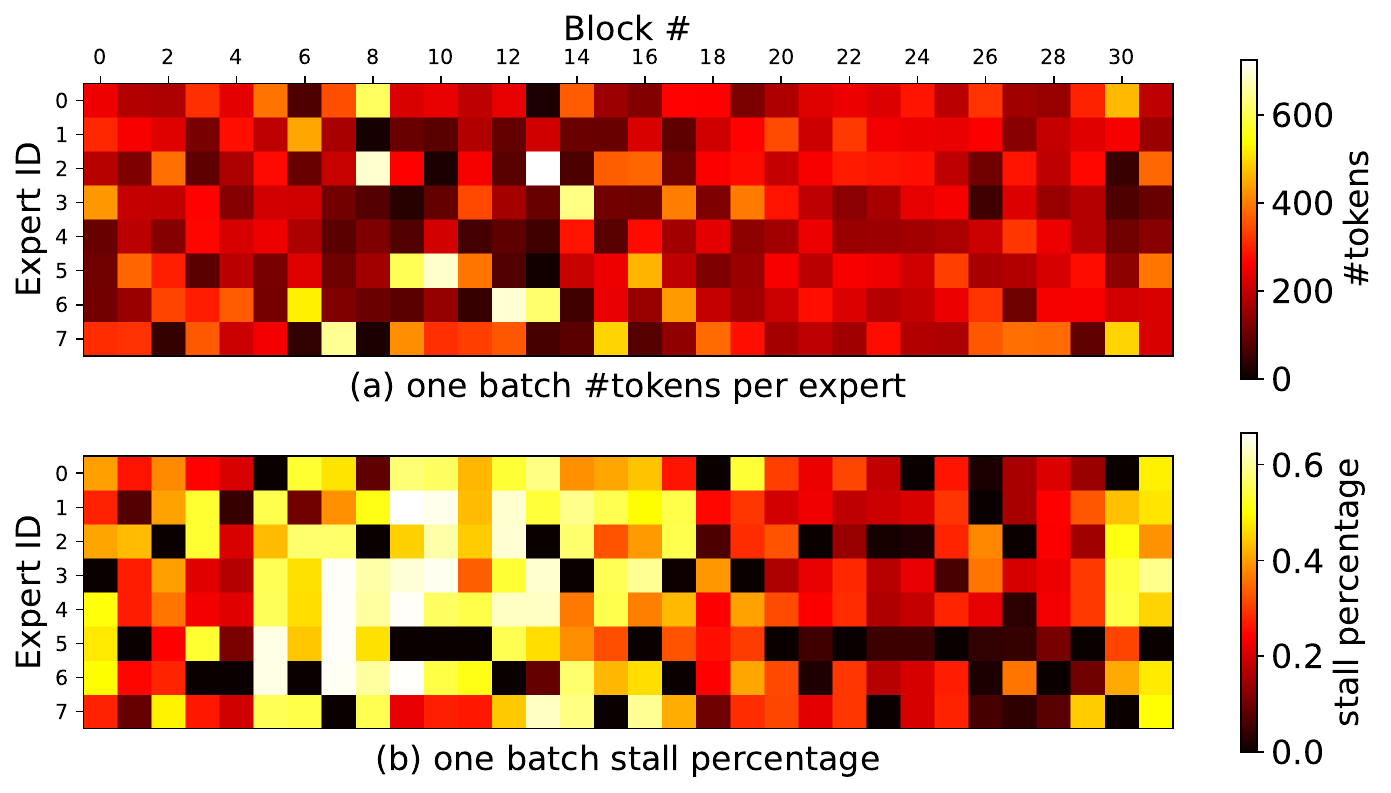}
    \caption{(a) expert load skew of a single iteration and (b) resulting GPU stall time fraction while serving Mixtral 8x7B with databricks-dolly-15k dataset at 100 req/s input rate on DGX A100 40GB (8x A100 40G with NVSwitch) using SGLang with expert parallelism. Mixtral 8x7B has 32 decoding blocks and 8 experts per block.}
    \label{fig:skewAndStall}
\end{figure}

Second, the group communication required by EP (all-to-all or all-gather operations to exchange tokens before and results after the parallel expert layers) \textbf{introduces significant stalling when load skew is present.} Experts receiving many tokens ("hot experts") take considerably longer to complete their computation. This creates a "straggler" effect, where all other devices must wait idly for the device hosting the hottest expert. This waiting time directly translates to lost GPU utilization, sometimes accounting up to 70\% of GPU time in skewed scenarios (see Figure~\ref{fig:skewAndStall}). Crucially, simply increasing the overall batch size to improve the per-expert computational efficiency (addressing the first challenge) can exacerbate this straggler problem, as it increases the execution time variance between the hottest and coldest experts.

Faced with these efficiency challenges inherent to EP under load skew, developers of prominent recent MoE models such as Mixtral~\cite{mixtral}, Grok~\cite{grok}, and DBRX~\cite{dbrx} have often employed Tensor Parallelism (TP) instead of EP for the expert layers. In TP, each expert's parameters are sharded across all participating GPUs. Since every GPU processes a piece of every expert, the computational load is perfectly balanced across devices, eliminating the straggler problem caused by uneven token distribution. However, TP introduces its own significant communication overheads, requiring frequent and high-volume data exchanges between GPUs for each expert computation. Furthermore, while TP balances load across GPUs, it may not fully resolve the computational inefficiency since cold experts still execute at small batch sizes. Primarily due to the high communication costs, TP-based MoE implementations struggle to scale efficiently beyond a single node (typically 8 GPUs connected via high-speed interconnects like NVLink). This limitation motivates the search for more scalable and efficient serving strategies for large MoE models. 



\subsection{Disaggregating Prefill from Decoding}
Disaggregating the prefill phase from decoding has become increasingly standard in large-scale LLM serving systems~\cite{distserve, loongserve, mooncake, hu2024infer, liang2025injecting, jin2024pdserve}. The reason is that prefill typically faces tighter time-to-first-token (TTFT) requirements and is often compute-bound, so it can take advantage of more aggressive parallelization (e.g., intra-operator parallelism) to achieve low latency. By contrast, decoding—especially when it must generate multiple tokens per request—tends to be more HBM-bandwidth-bound and exhibits smaller, frequent computational steps. This mismatch between the phases causes significant interference when they are colocated on the same GPU, making it harder to meet both TTFT and time-per-output-token (TPOT) targets simultaneously~\cite{distserve, sarathiserve}. Thus, by separating prefill onto its own GPUs, the system can tailor resource allocation and model parallelism strategies to precisely satisfy TTFT constraints, leaving the decoding side free to concurrently maximize throughput~\cite{distserve}. However, while prefill can be readily scaled up to utilize GPUs effectively, achieving good GPU efficiency for decoding is much more challenging, particularly in expert-parallel MoE architectures, where routing and load imbalance introduce extra complexity. Therefore, this paper concentrates on tackling the harder problem of high-throughput decoding, with a specific focus on optimizing expert parallelism in MoE-based LLMs.

\subsection{Trends of Efficient Attention Mechanism}
\label{s:trendAttn}

The attention mechanism's high memory bandwidth and capacity demands have made it a primary bottleneck for extending context length in LLMs. Consequently, numerous efforts focus on reducing its resource consumption.

A major trend involves evolving from standard Multi-Head Attention (MHA) \cite{vaswani2017attention} to variants like Grouped-Query Attention (GQA) \cite{ainslie2023gqa} and Multi-Query Attention (MQA) \cite{shazeer2019fast}. By sharing Key (K) and Value (V) projections across query heads (partially in GQA, fully in MQA), these methods substantially reduce the size of the memory-intensive KV cache. Other architectural ideas like Multi-Layer Attention (MLA) \cite{deepseek} explore cross-layer information processing. Complementary to architectural changes, KV cache quantization reduces the precision of stored K/V tensors (e.g., to INT8 or lower) \cite{yao2022zeroquant, xiao2023smoothquant, frantar2023gptq}, further decreasing memory footprint and bandwidth usage.

For extremely long contexts exceeding device memory, efficient KV cache management strategies have been suggested. Techniques like InfiniGen \cite{inifinigen} employ intelligent offloading and management to handle vast KV caches with bounded memory growth, mitigating the latency penalties of naive offloading to slower memory tiers. Furthermore, research explores specialized hardware, such as Processing-in-Memory (PIM) \cite{kwon2025lolpim}, which performs computations closer to memory to alleviate the data movement bottleneck inherent in attention.

As these diverse optimizations make attention more efficient, we anticipate that the primary performance bottleneck, particularly in Mixture-of-Experts (MoE) models, will shift towards the execution of the large expert layers.

\subsection{Our Approach: Asynchronous Expert-Parallel Decoding}
To address the efficiency challenges in large-scale MoE serving, we propose \textbf{asynchronous expert parallelism}, where each device can execute and forward output independently in an asynchronous manner. 
The key technique is layer-wise scheduling: queuing tokens at the granularity of individual layers~(which we call $\mu$-queuing) and adaptively re-batching and executing just in time with the tokens so far accumulated at the layer's own $\mu$-queue. 
Due to adaptive re-batching, GPUs do not need to wait for barrier-style all-to-all communication to finish. Instead, they stay busy as long as enough load is offered at any layer. 
By colocating more than one expert and layer on a GPU, scheduler can multiplex layers to prioritize execution of hot experts with enough input tokens and let cold experts to accumulate more tokens before execution. 

Despite the advantages offered by the layer-wise scheduling, three key challenges must be addressed to achieve optimal performance.

\begin{enumerate}
    \item Token-level dependencies need to be carefully handled to preserve the semantics of the Top-K gating function. This must be achieved while allowing tokens from different sequences to be processed independently, thereby maintaining efficiency.
    \item Having many layers that can be executed asynchronously can increase the queuing delay as tokens wait for their turn for execution. The layer placement and scheduling algorithm should be able to minimize queuing delays to ensure low latency and high throughput.
    \item An effective communication mechanism is required in replacement of all-to-all. This mechanism must facilitate the transfer of tokens between nodes without causing device stalls. 
\end{enumerate}

In the following section, we will discuss how we addressed these challenges.

\section{Design}

To demonstrate the benefits of asynchronous expert parallelism, we designed a prototype MoE inference serving system, \name. 

\subsection{Overview}

\begin{figure}
    \centering
    \includegraphics[width=0.9\linewidth]{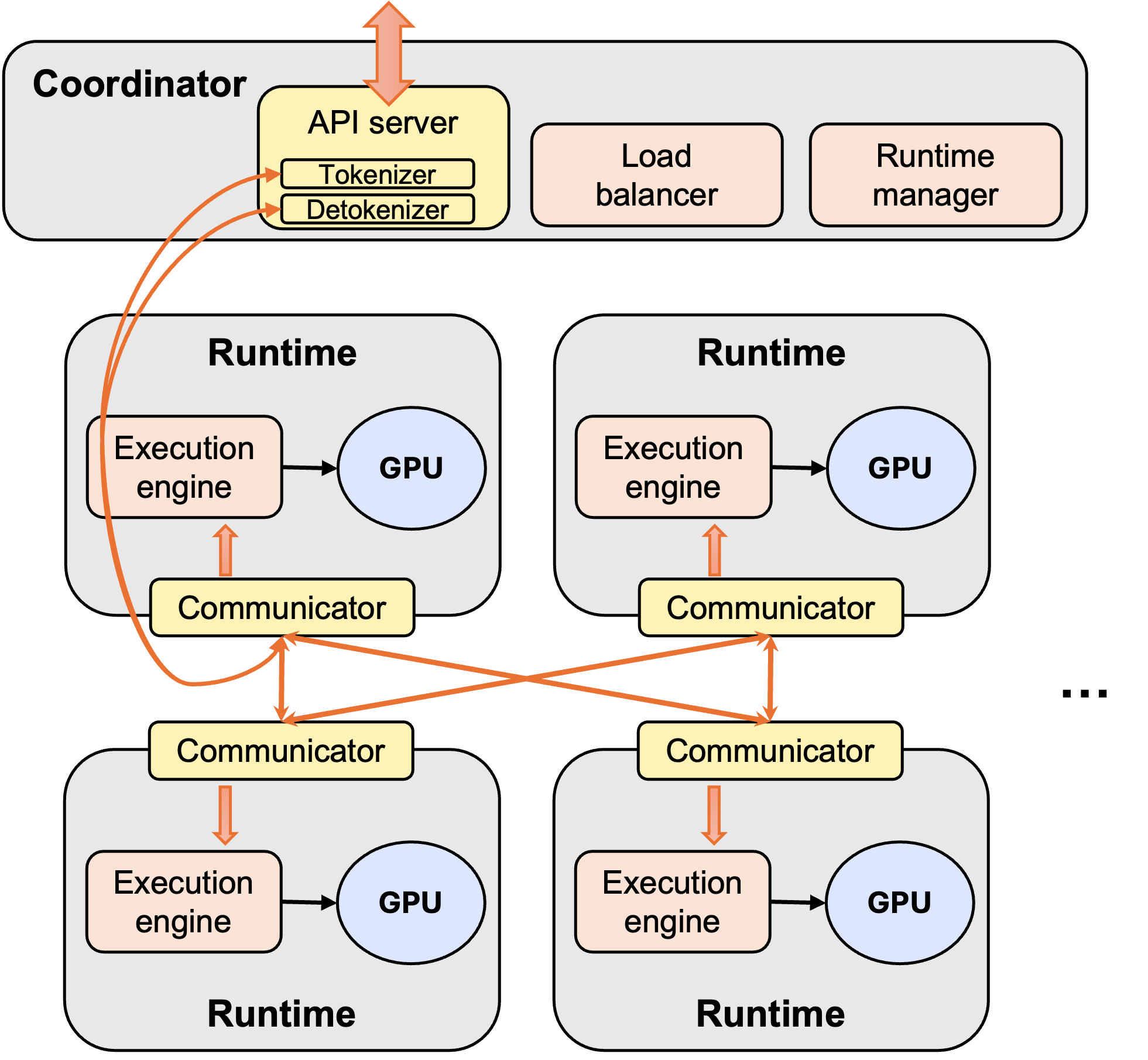}
    \caption{System architecture of \name. 
    }
    \label{fig:overview}
\end{figure}

\name is a Mixture-of-Experts (MoE) large language model (LLM) serving system compatible with vLLM~\cite{vllm}. \name splits MoE models into granular layers, allowing each layer to be individually scheduled asynchronously without being blocked by group communication across parallel GPUs. Specifically, we assume MoE architecture with multiple decoding blocks, each composed of an attention layer followed by a set of expert layers. We consider MoE's gating and top-K merge operators as part of the attention layer. Each decoding block is also often called one layer in some literature.

\name supports data parallelism (DP) for attention layers and expert parallelism (EP) for expert layers, which aligns with today's standard for MoE deployments~\cite{glam, deepseekv3, deepspeedmoe}. However, \name does not use blocking group communication before or after these parallel-execution layers.

As shown in Figure~\ref{fig:overview}, \name is composed of two types of services: coordinator and runtime. 
The coordinator is a CPU-run service, which may reside on the same host machine as GPUs (for small-scale deployments) or independently (for large-scale deployments). The coordinator consists of three components: API server, load balancer, and cluster manager. The API server handles incoming serving requests and maintains request states throughout the auto-regressive decoding process. 
It also includes a tokenizer and de-tokenizer. The load balancer monitors GPU memory usage across attention data-parallel (DP) ranks and assigns each new request to the rank with the most available memory. Once assigned, a request remains bound to the same DP rank for its entire auto-regressive decoding process, ensuring that all attention-related computation can reuse the key-value cache on the same GPU. The cluster manager oversees runtimes for GPUs, including setting up communication channels during initialization and tracking GPU memory usage, which is then provided to the load balancer. Together, these components serve as a centralized controller in \name to manage requests and workers.

Another component of \name is the runtime. \name assigns a separate runtime instance to each GPU. Each runtime handles the execution of several layers assigned by the coordinator. The runtime receives tokens from another runtime or tokenizer in the API server (for new requests), executes the appropriate layer for each token, and forwards tokens to either another runtime or the API server (for completed requests). The runtime manages layer-wise token queuing with dependency tracking (\S\ref{s:exec_engine}), GPU task scheduling (\S\ref{s:defrag_scheduler}), and efficient communication between runtimes (\S\ref{s:comm}). To minimize overheads, each token's input/output tensor data are kept in GPU memory and transferred directly to another GPU. The runtime manages these tensor data with metadata on the CPU side.

\subsection{Execution engine}
\label{s:exec_engine}

The core of the runtime is the execution engine, which asynchronously accumulates tokens and executes layers with the correct input on its GPU. Unlike typical MoE serving systems that rely on group communication, \name allows flexible asynchronous executions, which may reorder tokens in a random fashion throughout an iteration. Therefore, \name directly manages incoming tokens by associating each token's tensor data with metadata. By referring to this metadata, the execution engine can select the optimal layer for execution and supply the correct input tensor data.

\begin{table}
  \centering
  \small
  \begin{tabular}{|p{0.9\columnwidth}|}
    \hline
    \textbf{Metadata for Tokens} \\ \hline
    \begin{itemize}
      \item \texttt{RequestID}: used for tracking generated output token and attention DP rank
      \item \texttt{LayerID}: indicates the layer this token should be used as input for. It is composed of <block\#> + <expert\#> or <attn DP rank>
      \item \texttt{Tensors[]}: reference to input tensors on the GPU
      \item \texttt{Prefill\_length}: used for attention
      \item \texttt{Topk\_weights}: used for top-k token merging
    \end{itemize} \\ \hline
  \end{tabular}
  \caption{List of token metadata items}
  \label{tab:token_metadata}
\end{table}

Execution engines retain input tensor data on GPUs to avoid CPU-GPU data transfer overheads and track these tensor data with metadata on the CPU. Table~\ref{tab:token_metadata} lists the information tracked for each token. \texttt{RequestID} accompanies each token throughout the auto-regressive decoding process, allowing us to associate the generated output token with the user request. This is important since tokens may shuffle around due to asynchronous queuing and execution, making it impossible to infer each token's request ID based solely on its index in the global batch. Similarly, \texttt{LayerID} is set before forwarding a layer's output to the next layer, indicating the target layer for execution with this token. If the target layer is an expert layer, \texttt{LayerID} is composed of \texttt{<block\#>} + \texttt{<expert\#>}. If it is an attention layer, \texttt{LayerID} comprises \texttt{<block\#>} + \texttt{<attn DP rank>}. \texttt{LayerID} is first used by the communicator to determine the next destination of the token and subsequently used by the receptor to queue the received token into the correct queue. This metadata structure retains references to input tensor data on the GPU (\texttt{Tensors[]}), which can be more than one if the target layer requires multiple input tensors. In addition to these three token management metadata items, a token may also carry additional metadata required for attention execution or MoE's top-K token merging.

\begin{figure}
    \centering
    \includegraphics[width=\linewidth]{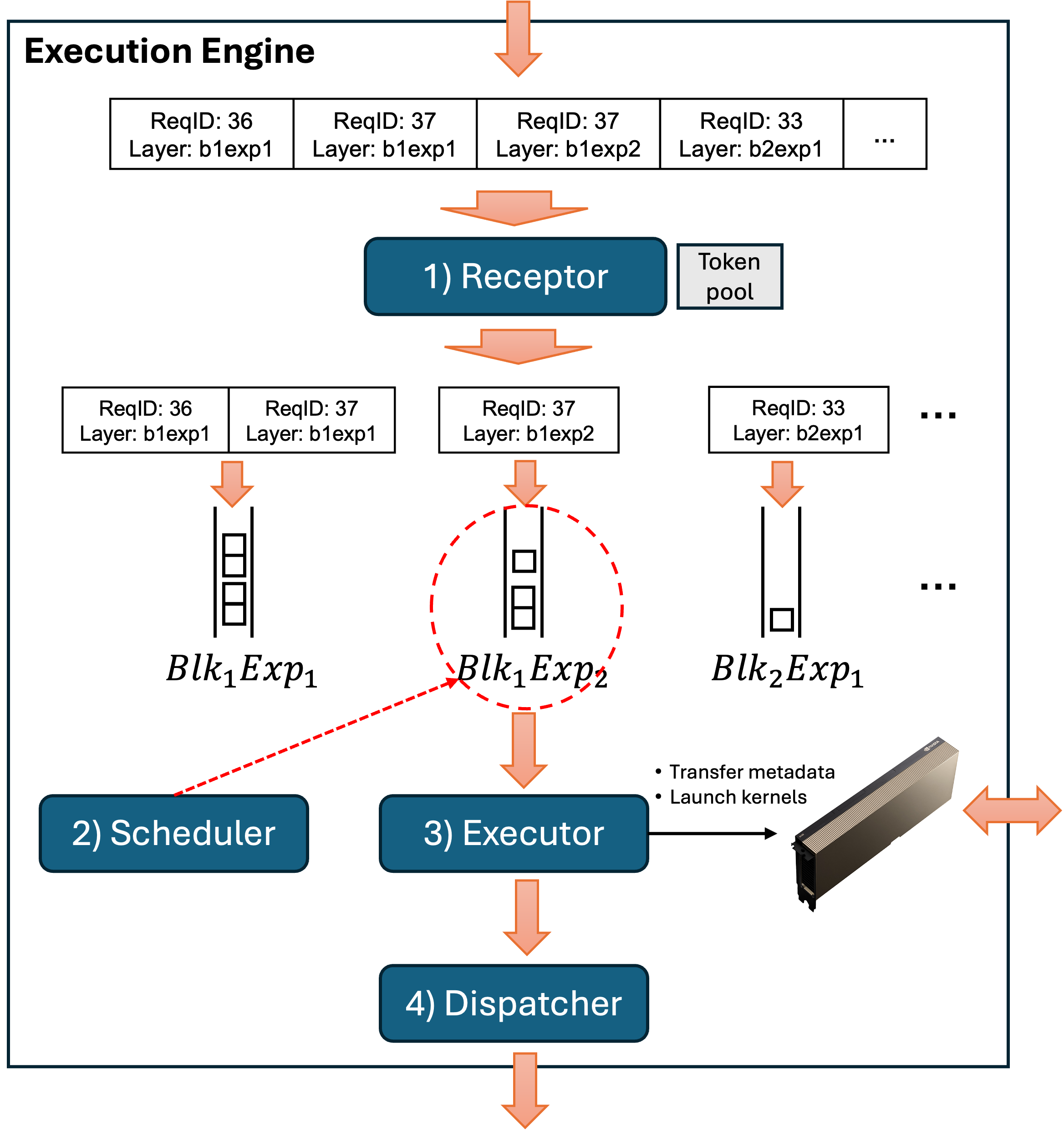}
    \caption{The data flow of a token batch within runtime involves four steps: (1) receptor puts incoming tokens into the corresponding layer-specific $\mu$-queues, (2) scheduler picks the optimal layer for execution, (3) executor runs the selected layer's computation on GPU, and (4) dispatcher assigns next destination for tokens
    }
    \label{fig:engine}
\end{figure}

Using the metadata, the execution engine processes input tokens in 4 stages as shown in Figure~\ref{fig:engine}. 
First, receptor is the entry point for incoming token batches fetched from the communicator. Receptor segregates the received metadata of tokens by the \texttt{LayerID} and enqueue them to the corresponding layers' queues. 
Second, whenever GPU becomes idle, scheduler picks the layer whose queue should be drained for execution. Third, executor drains the selected queue, transfer necessary metadata to the GPU (e.g., prefill length for attention layers), and launch corresponding kernels for execution. Here, executor launches our custom CUDA kernel for preparing a contiguous input token batch from many individually arrived token batches. 
Lastly, when execution on GPU is finished, dispatcher re-labels the output tokens with next \texttt{LayerID}s, which are then forwarded to the communicator.

\parab{Top-K support:}
Supporting top-K ($K > 1$) requires additional mechanisms beyond those described above.
After routing (the last operator in the attention layer in \name), a token is duplicated $K$ times and dispatched to $K$ different experts for processing by the dispatcher.  
These duplicated tokens serve as input tensors for the attention of the next block (whose first operator is the top-K merge operator). Until all inputs are ready, we cannot execute the attention layer. To ensure the scheduler and executor are dealing only with ``ready'' tokens whose inputs are already on the GPU, the receptor retains the incomplete tokens until all input tensors have arrived and are ready. When a new token batch arrives, the receptor inspects whether a token is ready by itself. If the token needs more than one input for the next layer execution, it holds the token in a \textit{token pool}. At the token pool, previously duplicated tokens (identified by the tuple \texttt{<RequestID, LayerID>}) are merged into a single token. Once a token is merged and has all input tensors ready, the receptor moves the token to the corresponding $\mu$-queue for scheduling.

\parab{Executor details:} 
Executor directly controls the GPU. Once the layer to execute is selected by scheduler, it performs forward computation for the selected layer. 
KV cache management is also handled by executor. \name leverages paged attention\cite{vllm}, and each executor manages its GPU's block table that maps requests to key-value~(KV) cache blocks. Although layers operate independently, all attention layers on the same GPU share one page table as KV cache is are isolated for each layer. A new KV slot is allocated for a token only upon entering the first layer, allowing block table reuse across layers and reducing allocation overhead. While all layers follow the same execution model, the first attention layer requires additional processing to convert input tokens into embeddings. 

Furthermore, the runtime with the first attention layer includes a sampler, which sample out previous iteration's output tokens from embeddings. We place a separate sampler on each GPU with the first attention layer to avoid extra communication overhead. The sampler is treated equivalently to other attention layers and must be scheduled before execution.

\parab{Dispatcher details:} After each execution, the output tensors should be forwarded to the another runtime and GPU that are in charge of the next layer of the model. The dispatcher coordinates this output forwarding process. After attention layer execution (thus, next is an expert layer), dispatcher identifies each token’s assigned expert and permutes tokens by expert ID to group them. The permuted tokens are then split into several smaller batches and sent to appropriate expert workers over the network based on expert placement. In expert workers, tokens are instead permuted by their assigned attention DP rank, as their context remains on the same attention worker. The dispatcher also increases the layer ID of each batch by one, reflecting their transition to the next layer’s attention module.

\subsection{Layer placement}

\name's default placement strategy disaggregates attention layers from expert layers and colocates all layers of each type across all decoding blocks. For example, a GPU/runtime handling Expert 1 will host all Expert 1 layers across all decoding blocks.

We colocate each layer type across all decoding blocks for several reasons, as done by many other MoE systems~\cite{glam, deepseekv3, deepspeedmoe}. An alternative placement strategy would be sharding models across decoding blocks, essentially forming pipeline parallelism (PP). We chose not to shard models into multiple pipeline stages because pipelining can cause load imbalance and result in high latency. Instead of PP, allocating GPUs for more data parallelism (DP) for attention or expert parallelism (EP) for experts can reduce iteration time. While PP may reduce queuing delay for the first token decoding, it introduces higher inter-token latency, which is more significant for auto-regressive decoding. However, when there are abundant GPUs, \name can utilize those GPUs to reduce collocation and form multiple pipeline stages.

One of the key concerns with colocating multiple layers onto a single GPU/runtime is queuing delay. By colocating across all blocks, we can optimize scheduling by exploiting the precedence and ordering of layers (\S\ref{s:defrag_scheduler}). Since tokens proceed to higher block\#, the scheduler can try to congregate most tokens to one or two consecutive blocks, minimizing queuing delay for the majority of tokens.

We chose to disaggregate attention from experts to enable further layer-type-specific optimizations. Disaggregated deployment allows us to use a different number of GPUs for attention and expert layers. We observed that for some longer context generation tasks, KV-cache space in the GPU memory limits the number of concurrent requests in the decoding process, leading to GPU under-utilization during expert layer computation. By allocating more GPUs for attention, we can achieve higher throughput without needing additional GPUs for experts. Additionally, due to the memory capacity and bandwidth-intensive nature of attention layers, there are increasing efforts to adopt heterogeneous hardware for attention layers. \name can naturally adopt this improved hardware for attention. Finally, by disaggregating, we can better understand the benefits of Asynchronous Expert Parallelism in improving the efficiency of expert computation.

We considered developing a placement optimizer for a given cluster setting, but it is beyond our focus on validating the benefits of AEP. However, \name opens up opportunities for more flexible placement optimization, especially in heterogeneous clusters.

\subsection{Defragging Scheduler}
\label{s:defrag_scheduler}

Each runtime in \name hosts multiple layers. Whenever the GPU becomes available, the scheduler selects one of the layers for the next execution. We found that the selection strategy can significantly impact the latency and throughput of \name.

\begin{figure}
    \centering
    \includegraphics[width=\linewidth]{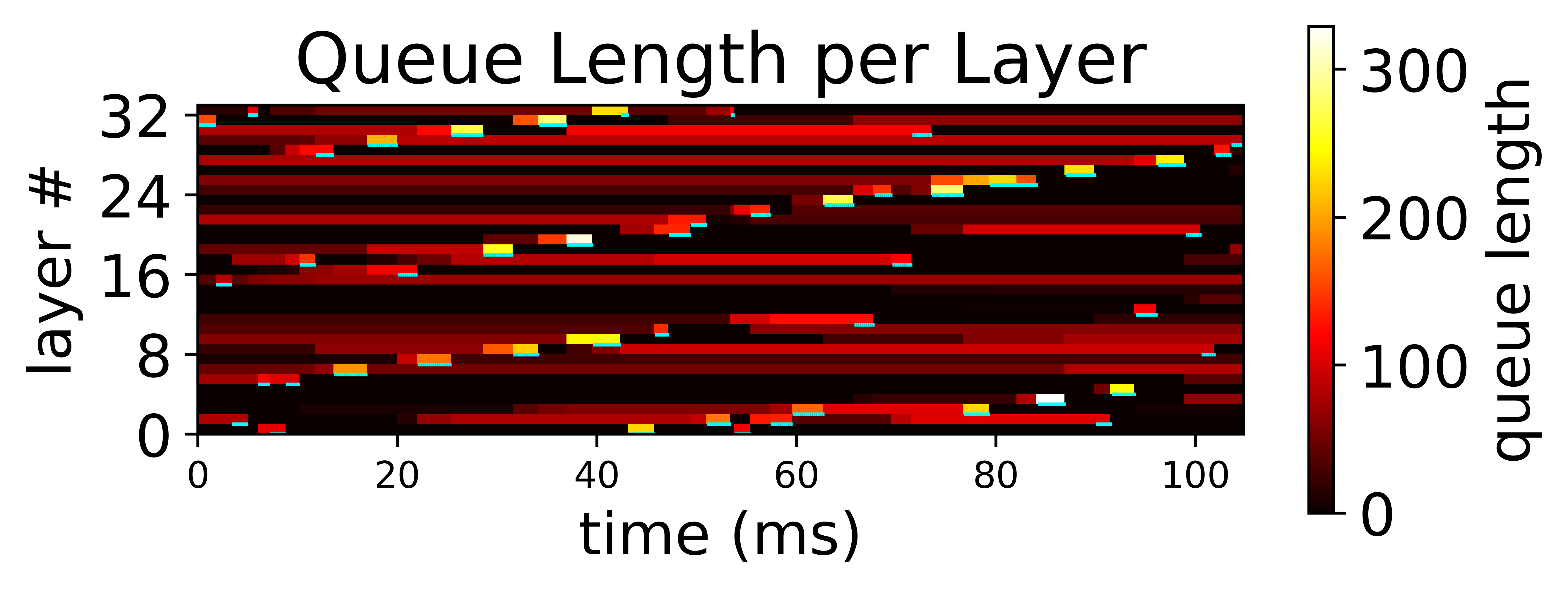}
    \caption{$\mu$-queue depth of an attention runtime with MTFS scheduling. Input rate is 200 requests for each second. Trace is collected from Figure \ref{fig:scheduler-ablation-load80} top-2 setting.}
    \label{fig:sched-mtfs}
\end{figure}

There are two strawman strategies that motivated \name's defragging scheduler. The first strawman is selecting the layer with the most tokens in the queue, which we call the ``\textit{most-token-first-serve} (MTFS)'' strategy. At first glance, prioritizing the layer with the most tokens over layers with fewer tokens sounds reasonable, as it can reduce queuing delay for more tokens. However, this strategy causes an interesting problem: batch fragmentation. Figure~\ref{fig:sched-mtfs} shows an example of the queue depth of each layer and their executions over time. In \name, a batch of tokens is distributed to many GPUs for attention DP. Then, some attention GPUs will return their attention output earlier than others. Conversely, the attention output batch is split and distributed to many expert GPUs. Because \name doesn't rely on blocking group communication, which merges all messages from all nodes, token batches naturally get fragmented unless they sit in the queue long enough to wait for more tokens. The \textit{most-token-first-serve} strategy doesn't help here. It tends to leave out the last slice of tokens for a layer since the next layer may accumulate more tokens by then. Every layer in the execution pipeline will leave out these orphans, leading to disorganized and fragmented batches. This is not ideal, as it increases queuing delay and lowers execution efficiency.

The opposite strawman is the \textit{first-layer-first-serve} (FLFS) strategy, which prioritizes earlier layers (e.g., lower block\#). By strictly prioritizing any earlier layer with some tokens, FLFS aggressively defragments execution batches and tries to maintain all tokens within one frontier layer. This extreme strategy performs reasonably well thanks to the autoregressive decoding; a well-merged wave of tokens will benefit the next iteration as well. However, it still has some drawbacks. Any newly introduced tokens (new requests) will take priority until they are merged into the main wave of tokens. With many short requests, the system may live lock and can hardly finish any requests. A better behavior would be for these tokens to wait until the main wave picks them up in the next iteration.

\begin{algorithm}[t]
\small
\caption{Defragging Scheduler}\label{alg:defragging_scheduler}
\begin{algorithmic}[1] 
    \State \textbf{Input:} $N_B$: NumBlocks, $N_E$: NumExperts, $Q[l, g]$: TokensInQueue, $\delta$: WeightDecay
    \State \textbf{Output:} $(b^*, e^*)$: Optimal (block, expert) to schedule
    \State Init $Scores[N_B][N_E] \gets 0$

    \For{$b \gets 0$ to $N_B - 1$}
        \State $LScore \gets 0$ \Comment{Calculate lookahead score}

        \For{$k \gets 1$ to $K$}
            \State $b' \gets (b + k) \bmod N_B$
            \State $TotalTokens \gets \sum_{e'=0}^{N_E-1} Q[b'][e']$
            \State $LScore \gets LScore + \left(\frac{TotalTokens}{N_e}\right) \times \delta^k$
        \EndFor

        \For{$e \gets 0$ to $N_E - 1$} 
            \Comment{Add lookahead with \#tokens}
            \If{$Q[b][e] > 0$}
                \State $Scores[b][e] \gets LScore + Q[b][e]$
            \EndIf
        \EndFor
    \EndFor

    \State $(b^*, e^*) \gets \arg\max_{b, e} \; Scores[b][e]$ 
           \Comment{Pick layer with max score}
    \State \Return $(b^*, e^*)$
\end{algorithmic}
\end{algorithm}

From these two strawman strategies, our scheduling algorithm aims to achieve a balance: promoting defragmentation like FLFS while considering queue occupancy like MTFS, thereby preventing both excessive fragmentation and interruption by new requests. Algorithm~\ref{alg:defragging_scheduler} presents a simplified version of our defragging scheduler algorithm. It calculates a score for each layer by combining the number of tokens currently waiting in that specific queue with a weighted lookahead score. The lookahead score estimates the density of tokens in subsequent layers down the pipeline, decaying the contribution of farther layers. By incorporating this lookahead, the scheduler favors executing layers that precede a dense wave of upcoming tokens, encouraging consolidation and forward progress. Unlike FLFS, it still can still leave out some fragments to limit inefficient small-batch executions.

This combined approach allows the scheduler to dynamically adapt, processing larger, consolidated batches when possible while still efficiently managing the flow of tokens across all layers and mitigating excessive queuing delays.


\subsection{Communicator}
\label{s:comm}

Each runtime in \name includes a communicator module that manages point-to-point (P2P) communication across runtimes. To take advantage of the fast GPU interconnect and avoid PCIe bottlenecks, \name uses NCCL as the primary transport. However, there are two challenges in using NCCL: sender/receiver synchronization and variable tensor sizes.

\begin{figure}[h]
    \centering
    \includegraphics[width=\linewidth]{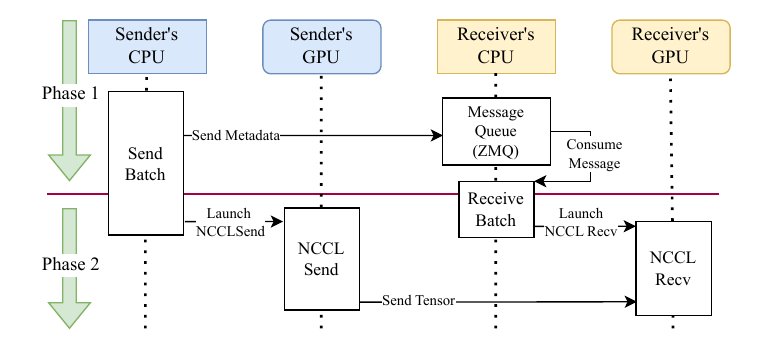}
    \caption{The communication in \name to transfer one batch between the sender and the receiver}
    \label{fig:communication}
\end{figure}

With NCCL's P2P API, both the sender and receiver GPUs must invoke the \texttt{NCCLSend} and \texttt{NCCLRecv} kernels simultaneously to initiate data transmission. Additionally, the receiver must know the sender GPU rank and tensor size in advance. In \name, due to varying user requests and scheduling, a runtime may receive batches with a dynamic number of tokens from many other GPUs at any time. Therefore, a mechanism is needed to set up this transfer before initiating NCCL.

Figure~\ref{fig:communication} illustrates our solution: a two-phase communication process. Before initiating NCCL transmission (Phase 2), the sender sends metadata to the receiver through a message queue library, ZeroMQ~\cite{zeromq}, on the CPU (Phase 1).
Each communicator maintains a message queue to iteratively consume metadata from any senders. Upon receiving new transmission metadata, it creates a NCCL buffer of the size specified in the metadata.

After exchanging metadata, NCCL transmission (phase 2) begins. The sender and receiver launch \texttt{NCCLSend} and \texttt{NCCLRecv} respectively, where the CPU side initiating the NCCL kernels on GPU streams. After launching the NCCL kernels, the CPU side proceeds to the next transmission task (such as checking the ZeroMQ queue) without waiting for the NCCL kernels to finish. Before releasing the received tensor to the scheduler, the communicator synchronizes with the GPU to ensure that the NCCL transmission is complete. The sender does not need to synchronize since the batch is no longer used after transferring. Consequently, a single-threaded communicator can send or receive multiple batches concurrently.

\section{Implementation}



We implement \name from scratch, comprising 6K lines of Python and 4.8K lines of C++ code. The runtime is primarily written in Python to take advantage of the highly optimized model execution infrastructure provided by vLLM \cite{vllm}. While the model executor resides in Python, the communicator, receptor, scheduler, and dispatcher of the execution engine are developed in C++ to reduce the overhead associated with layer-wise scheduling. These components expose interfaces to Python through pybind11. The scheduler and executor execute within the main Python thread, while the receptor and dispatcher run on dedicated POSIX backend threads to facilitate efficient communication-computation overlap. The usage of C++ helps to bypass python's Global Interpreter Lock~(GIL) and make all component operate concurrently.


\textbf{CUDA Graphs.} We build CUDA Graphs with pytorch to reduce the launching overheads of multiple small CUDA kernels for small batch sizes in the attention engine. Conventionally, serving systems record $G$ graphs for the entire model, each corresponding to a disjoint batch size range. In \name, however, scheduling and execution are performed at the layer level. We record $G$ graphs for every layer, leading to $L\times G$ layer-wise graphs, where $L$ denotes the number of layers. It incurs huge memory footprints as each graph maintains static data buffers. We alleviate the memory pressure by sharing the input buffer across all graphs. Meanwhile, intermediate and output buffers are allocated in the runtime, we track these buffers to get the results of computation.

Although CUDA Graph can effectively accelerate the attention engine when batch size is small, they are less beneficial for expert computation in \name. Execution in the expert engine is dominated by heavy GEMM~(General Matrix Multiplication) kernels, interleaved with a few lightweight kernels. The CPU is able to asynchronously launch kernels during the first GEMM kernel, thereby minimizing kernel launch latency. In \name, we observe this kernel launch latency smaller than the combined overhead copying data to CUDA graph input buffers.


\textbf{Batch Management in Attention Executor.} The attention executor imposes additional requirements on the input batch. Specifically, it allocates new key-value slots for incoming tokens and locates their corresponding key-value pages. Key-value metadata along with the current decoding lengths of the associated requests are then transferred from CPU to GPU memory, where they are later consumed by the paged attention kernels. After attention computation, the expert indices and weights for each token must be copied back from GPU to CPU to allow the dispatcher to correctly route the tokens. This execution workflow involves many small memory transfers between CPU and GPU. To optimize performance, we fuse these small copies into larger batched transfers, thereby reducing kernel launch overheads. Additionally, data transfers are offloaded to a dedicated CUDA stream, ensuring that communication does not block the main execution thread. We introduce and analyze the details in \S\ref{section:execution-breakdown}.


\section{Evaluation}

Our evaluation aims to answer the following questions:
\begin{enumerate}
    \item Does \name provide better throughput and latency than the state-of-the-art expert-parallel serving system? 
    \item On what kinds of workloads does AEP have an advantage over EP? 
    \item Does AEP allow better scalability than TP or standard EP?
    \item How much does defragging scheduler help on throughput and latency? 
    \item How much overheads does layer-wise scheduling incur? 
\end{enumerate}

\begin{table}
\centering
\begin{small}
\begin{tabular}{r|l}
\toprule
GPU     & 8 $\times$ NVIDIA A100-SXM4-40GB \\
Interconnect &  NVSwitch (600 GB/s for each GPU) \\
Network &  4 $\times$ 100 Gbps Elastic Fabric Adapter~\cite{aws-efa}\\
Driver  & {CUDA 12.4, cuDNN: v9.1.0, NCCL 2.22.3} \\
CPU     & {2 $\times$ AMD EYPC 64 cores @ 1.5 GHz} \\
RAM     & {988 GB} \\
OS      & {Ubuntu 22.04.1 (Linux 6.8.0-1021-aws)} \\
\bottomrule
\end{tabular}
\caption{AWS P4 instance configuration.}
\label{tbl:hardware-aws}
\end{small}

\centering
\begin{small}
\begin{tabular}{r|l}
\toprule
GPU     & 8 $\times$ NVIDIA A100-SXM4-80GB \\
Interconnect &  NVSwitch (600 GB/s for each GPU) \\
Driver  & CUDA 12.8, NCCL 2.25.1 \\
CPU     & 2 $\times$ AMD EYPC 64 cores \\
RAM     & 1800 GB \\
OS      & Ubuntu 22.04 (Linux 6.8.0-52-generic) \\
\bottomrule
\end{tabular}
\caption{Lambda instance configuration.}
\label{tbl:hardware-lambda}
\end{small}
\end{table}

To answer the questions above, we measured the performance of \name and state of the art serving system, SGLang~\cite{sglang}. We modified SGLang to bypass the prefilling stage by populating KV cache with dummy data and focus solely on decoding, which is consistent with \name's setup. For most evaluations (\S\ref{s:eval-overall}, \S\ref{s:eval-scheduler}), we benchmarked the decoding performance of \name and SGLang with Mixtral 8x7B which has 8 experts on the hardware listed in Table~\ref{tbl:hardware-lambda}. To mimic the realistic expert load skew, we profiled expert load distribution using Dolly dataset\cite{DollyDataset} and fitted it to an exponential distribution. We replaced the routing layer in Mixtral 8x7B with our own routing which randomly selects experts based on the profiled exponential distribution.

For scalability benchmark (\S\ref{s:eval-scalability}), we mimicked the Llama-V4 by increasing the number of experts of Mixtral 8x7B to 16 and using top 1 routing. This 16 experts model is deployed to 2 instances of AWS p4dn machines (Table~\ref{tbl:hardware-aws}), totaling 16 GPUs over datacenter networking.\footnote{Lambda cluster in Table~\ref{tbl:hardware-lambda} was not designed for large scale inference and has very slow networking ($\sim$10 Gbps).} We also replaced the routing layer with the exponential distribution modeled by profiling Mixtral 8x7B\cite{mixtral} with Dolly dataset.

\begin{figure*}[t]
    \centering
    \begin{minipage}{0.32\textwidth}
        \includegraphics[width=\linewidth]{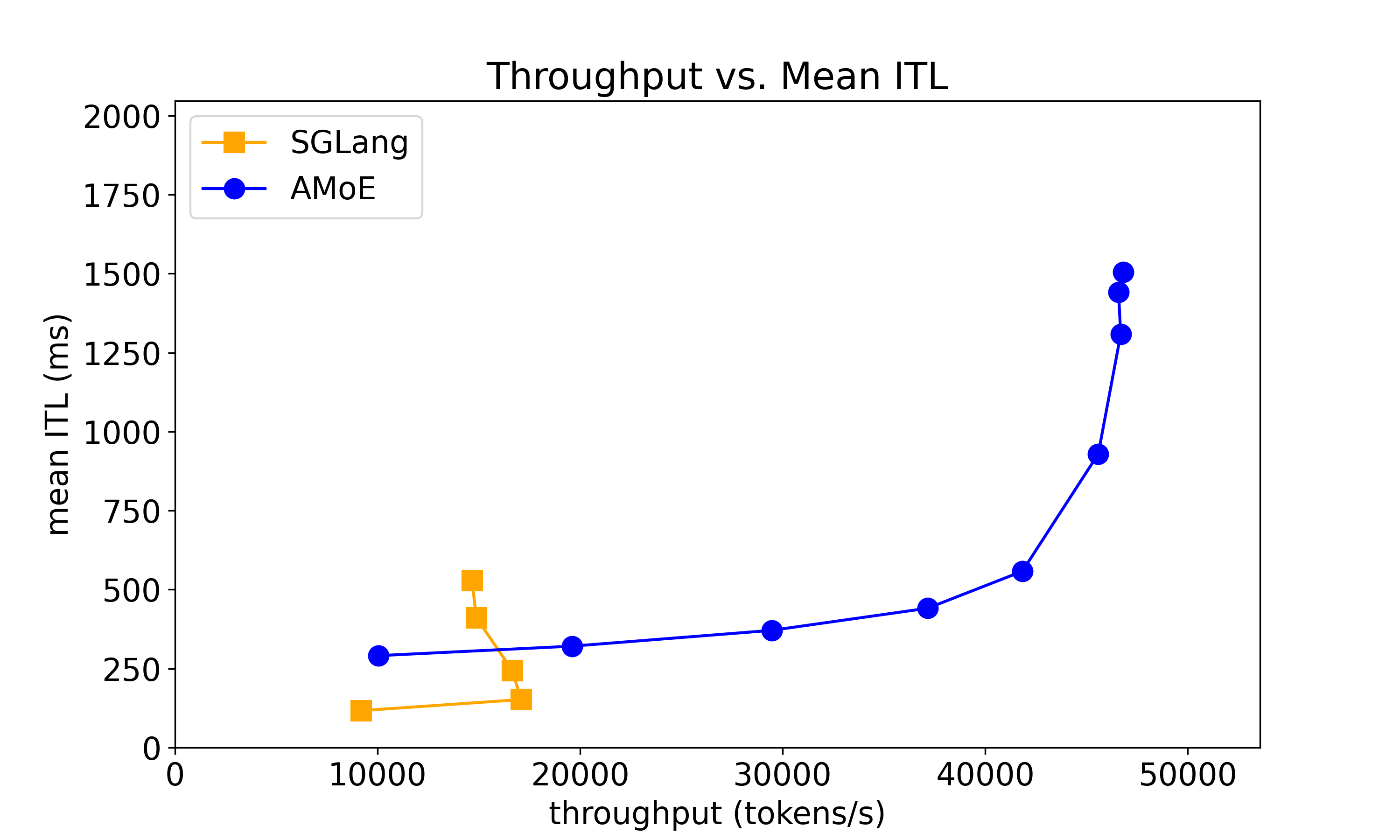}
        \subcaption{Top-1 with short workload}
    \end{minipage}
    \hfill
    \begin{minipage}{0.32\textwidth}
        \includegraphics[width=\linewidth]{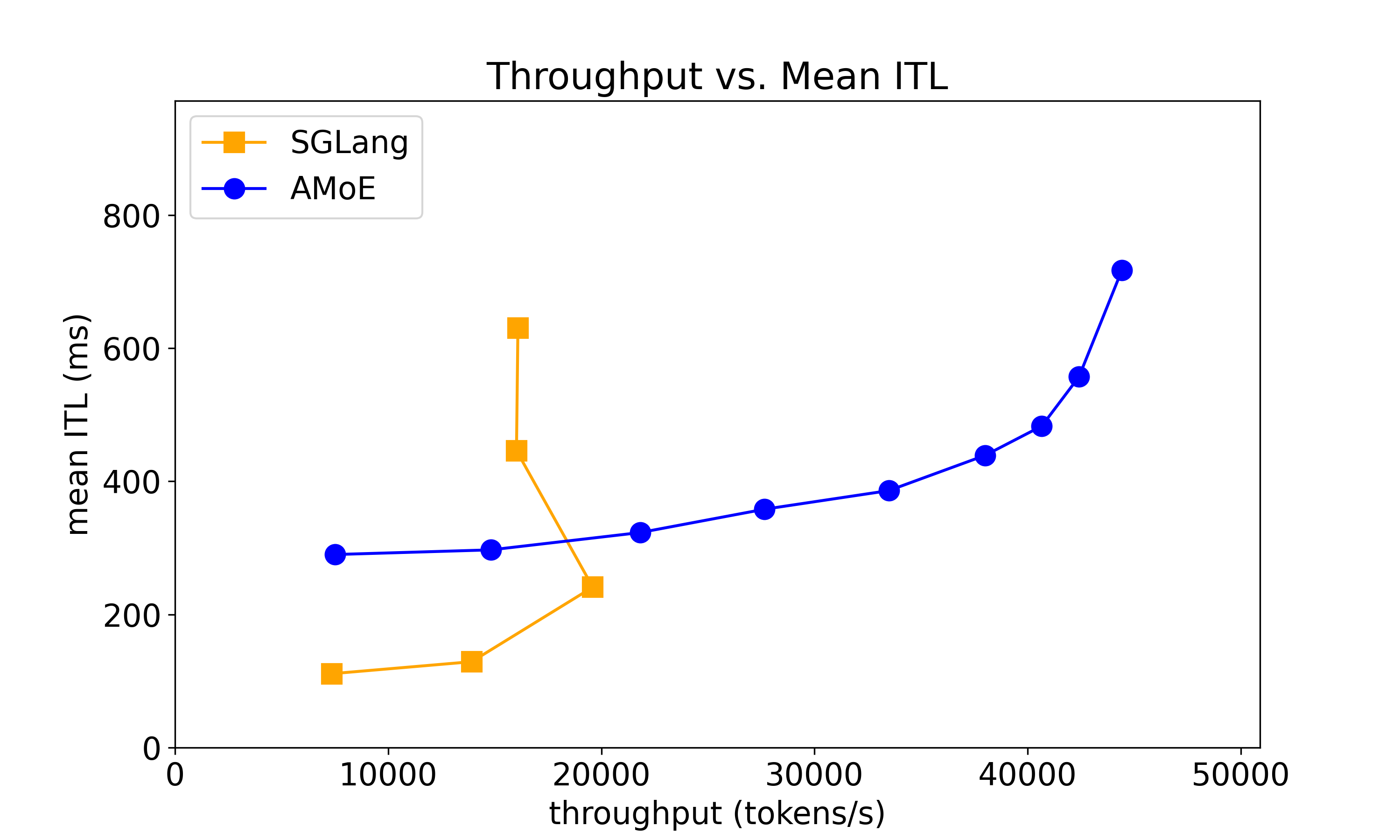}
        \subcaption{Top-1 with medium workload}
        \label{fig:mqa-top1-medium}
    \end{minipage}
    \hfill
    \begin{minipage}{0.32\textwidth}
        \includegraphics[width=\linewidth]{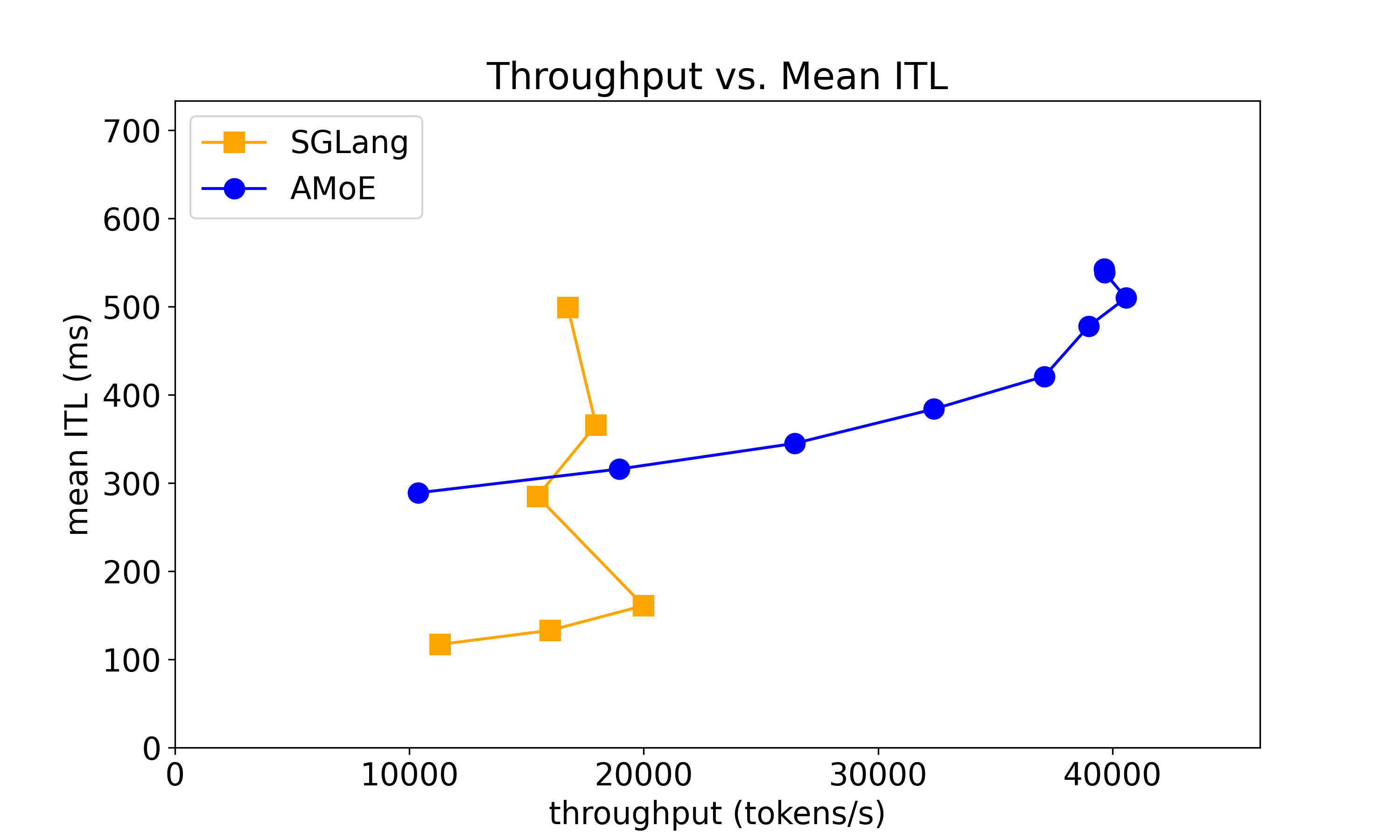}
        \subcaption{Top-1 with reasonable workload}
    \end{minipage}

    \vspace{0.4cm} 

    \begin{minipage}{0.32\textwidth}
        \includegraphics[width=\linewidth]{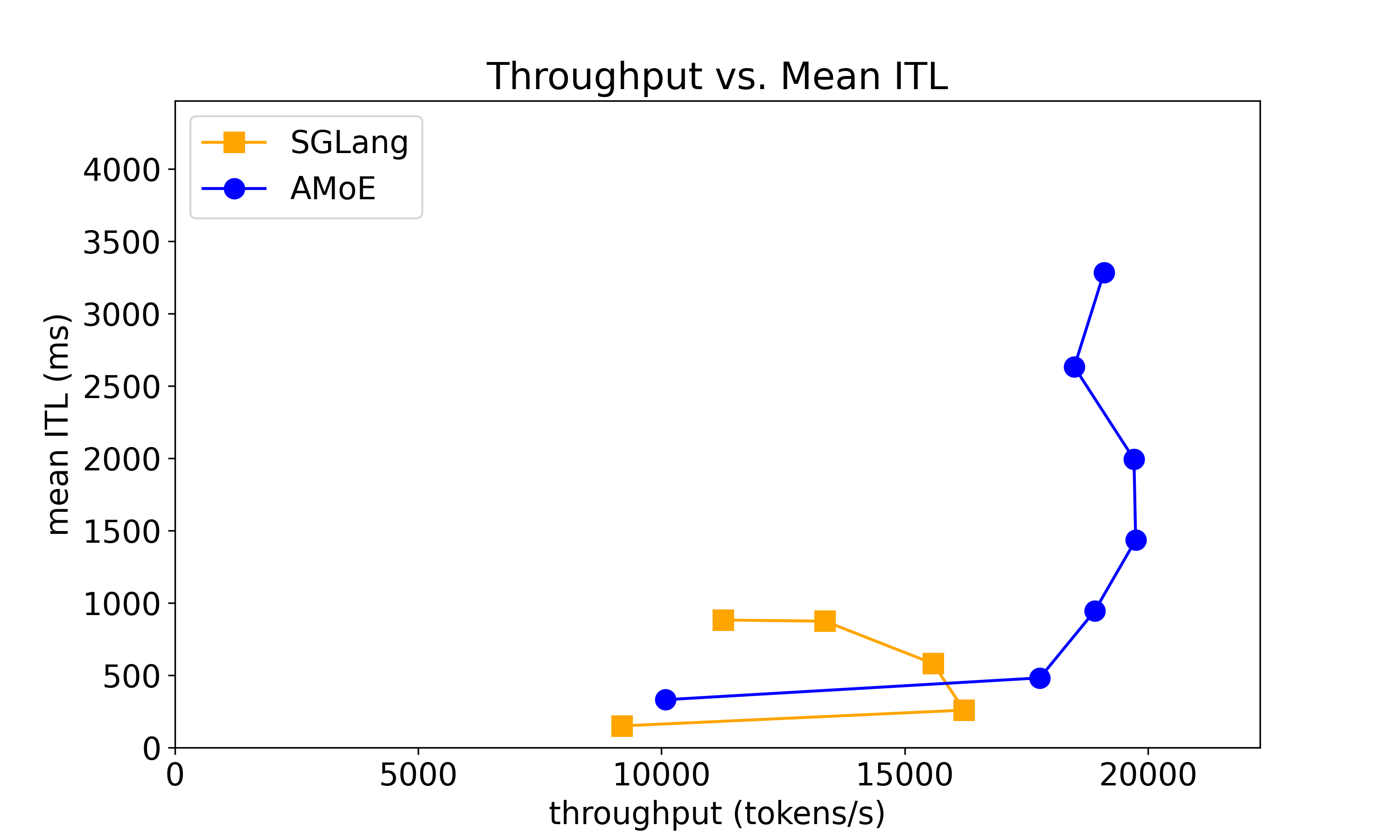}
        \subcaption{Top-2 with short workload}
    \end{minipage}
    \hfill
    \begin{minipage}{0.32\textwidth}
        \includegraphics[width=\linewidth]{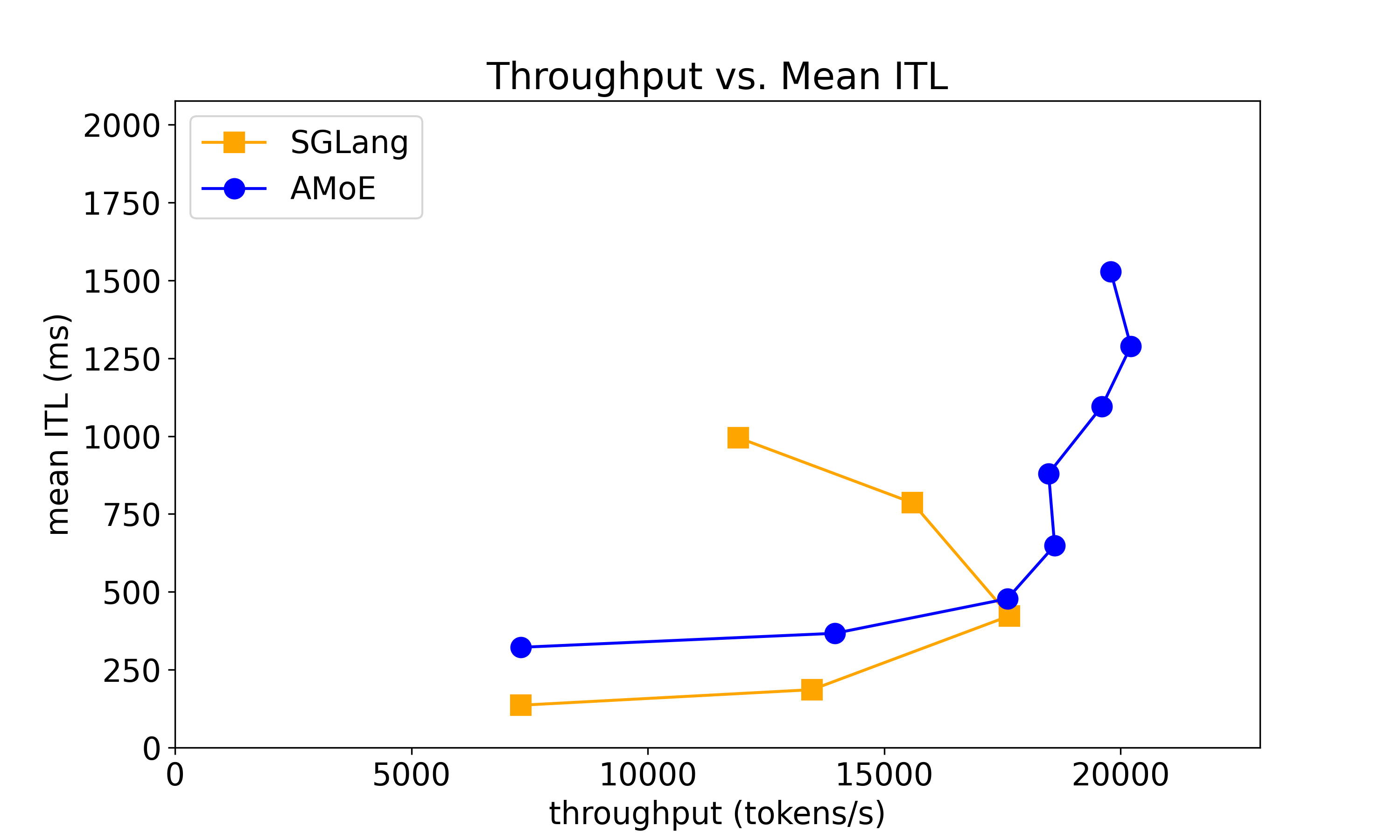}
        \subcaption{Top-2 with medium workload}
    \end{minipage}
    \hfill
    \begin{minipage}{0.32\textwidth}
        \includegraphics[width=\linewidth]{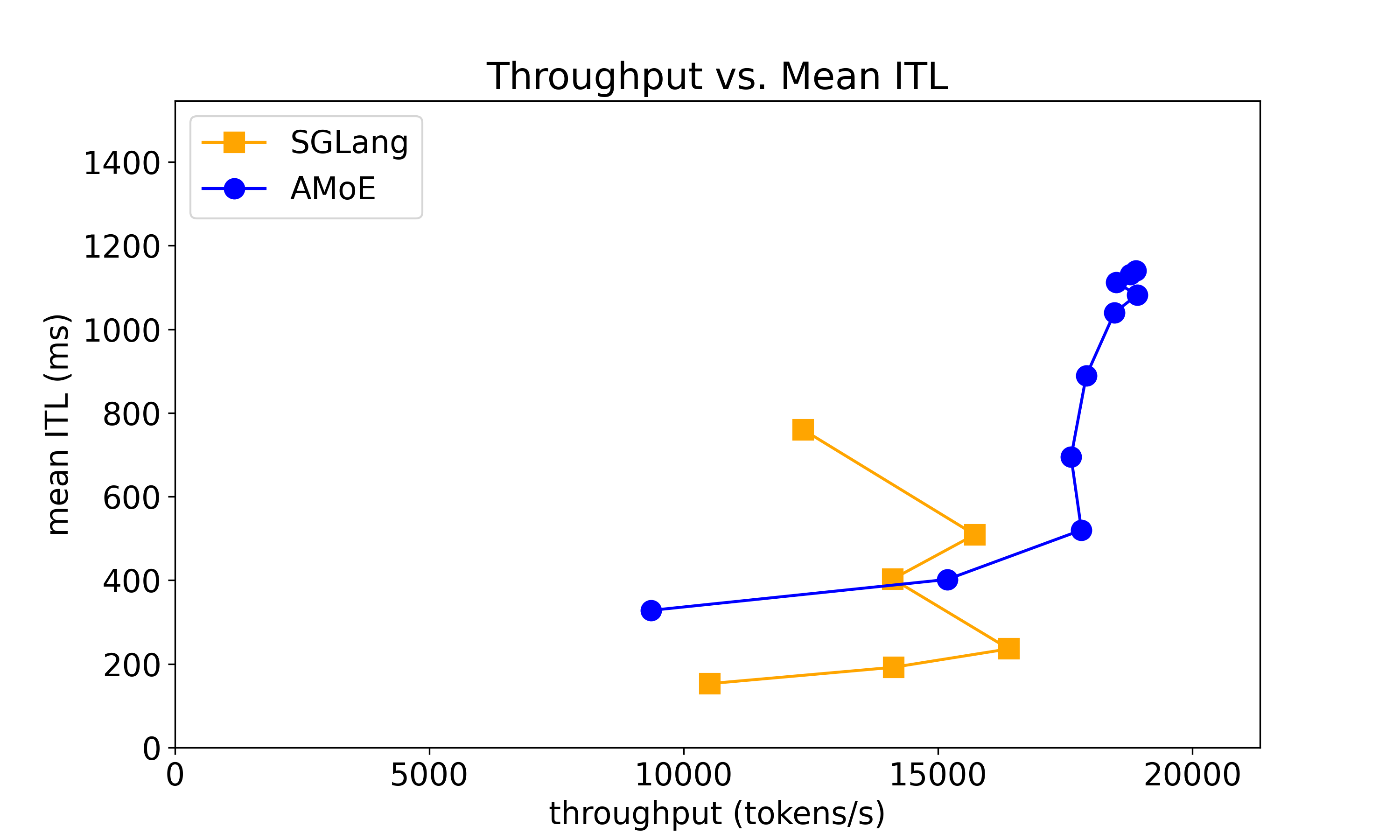}
        \subcaption{Top-2 with reasonable workload}
    \end{minipage}

    \caption{Overall performance comparison between \name and SGLang. Shows achievable throughput (x-axis) and corresponding observed inter-token latency (y-axis) under different routing and workload settings.}
    \label{fig:eval-overall}
\end{figure*}

\parab{Adjusting attention-expert intensity:} To emulate realistic, production-scale MoE serving workloads on our limited-scale hardware, we rebalance the load on attention and expert layers in two ways.

First, we modify Mixtral's attention mechanism from Grouped-Query Attention (GQA)~\cite{ainslie2023gqa} to Multi-Query Attention (MQA)~\cite{shazeer2019fast} to mitigate KV cache space limitations. With the original GQA, KV-cache capacity constraints both systems from processing many requests concurrently, resulting in roughly similar performance. As discussed in \S\ref{s:trendAttn}, there is a plethora of research on improving KV cache efficiency in attention mechanisms. Notably, DeepSeek proposes Multi-head Latent Attention (MLA)~\cite{deepseekv3}, which compresses KV cache usage by 10x without compromising model performance. Hence, we believe reducing KV cache contention via MQA is a reasonable approach to highlight AEP's advantage over standard EP for emerging MoE models.

Second, we maintain a balance between attention and expert computation intensity by using relatively short input and output token lengths. Based on DeepSeek and other production models, we believe expert computation is considerably more intensive than attention computation for large production-scale MoE models~\cite{deepseekv3, olmoe}. Unfortunately, due to our benchmark hardware limitations, we use a smaller version of Mixtral, namely Mixtral 8x7B, whose expert computation is about 2x lighter than the larger Mixtral 8x22B (Figure~\ref{fig:tputByBatch}). With long-context workloads, we observe that attention computation dominates in Mixtral 8x7B. Therefore, we focus on shorter decoding workloads so that the expert computation bottleneck is not obscured by the small MoE model.

\parab{Workloads:}
\label{s:eval-workloads}
We compare the performance of three different types of decoding workloads:
\begin{itemize}
    \item \textit{Short}: input [30, 70], output [70, 130]
    \item \textit{Medium}: input [50, 150], output [50, 250]
    \item \textit{Reasonable}: input [100, 300], output [100, 500]
\end{itemize}

For each workload, we generate new requests through a Poisson arrival process with a given rate, and each request picks its input length and output length with uniform random selection from the ranges listed above.

\subsection{Performance over various workloads}
\label{s:eval-overall}

We begin by comparing \name with our baseline serving system, SGLang, using Mixtral 8x7B with 8x A100 80GB GPUs on a single host. \name disaggregates attention layers to 4 GPUs with DP and uses the other 4 GPUs for expert layers with EP. SGLang uses DP for attention layers and EP for expert layers over all 8 GPUs.

Figure~\ref{fig:eval-overall} presents a high-level comparison of \name and our baseline SGLang under various routing (Top-1 and Top-2) and workload configurations. Each subplot in Figure~\ref{fig:eval-overall} plots the achievable token throughput on the x-axis against the corresponding average inter-token latency (ITL) on the y-axis. Across all scenarios in Figure 9, \name consistently achieves higher throughput than SGLang.

As depicted in Figures~\ref{fig:eval-overall} (a)–(c), \name consistently achieves higher throughput than SGLang when using Top-1 routing across all three request-length categories: 2.7x, 2.3x, and 2.0x for short, medium, and reasonable workloads correspondingly. We attribute this improvement to \name’s ability to dynamically re-batch tokens at each GPU and expert, thereby reducing GPU stalls even when the load distribution is skewed.
However, under low loads, SGLang shows lower ITL than \name. This is because of layer-wise scheduling overheads and attention disaggregation.

A similar trend persists under Top-2 routing, shown in Figures~\ref{fig:eval-overall} (d)–(f). \name retains its throughput advantage over SGLang, although the level of throughput improvement is less significant than Top-1. We suspect two reasons on this. First, Top-2 routing’s increased expert activation percentage (from 12.5\% to 25\%) partially tempers load skew by distributing tokens more evenly among the experts. Second, Top-2's token merge operation needs to wait for both outputs from experts, creating a partial synchronization point and reducing the benefit of asynchronous expert parallelism.

\subsection{Scalability (multi-node)}
\label{s:eval-scalability}

\begin{figure}[t]
    \centering
    \includegraphics[width=\linewidth]{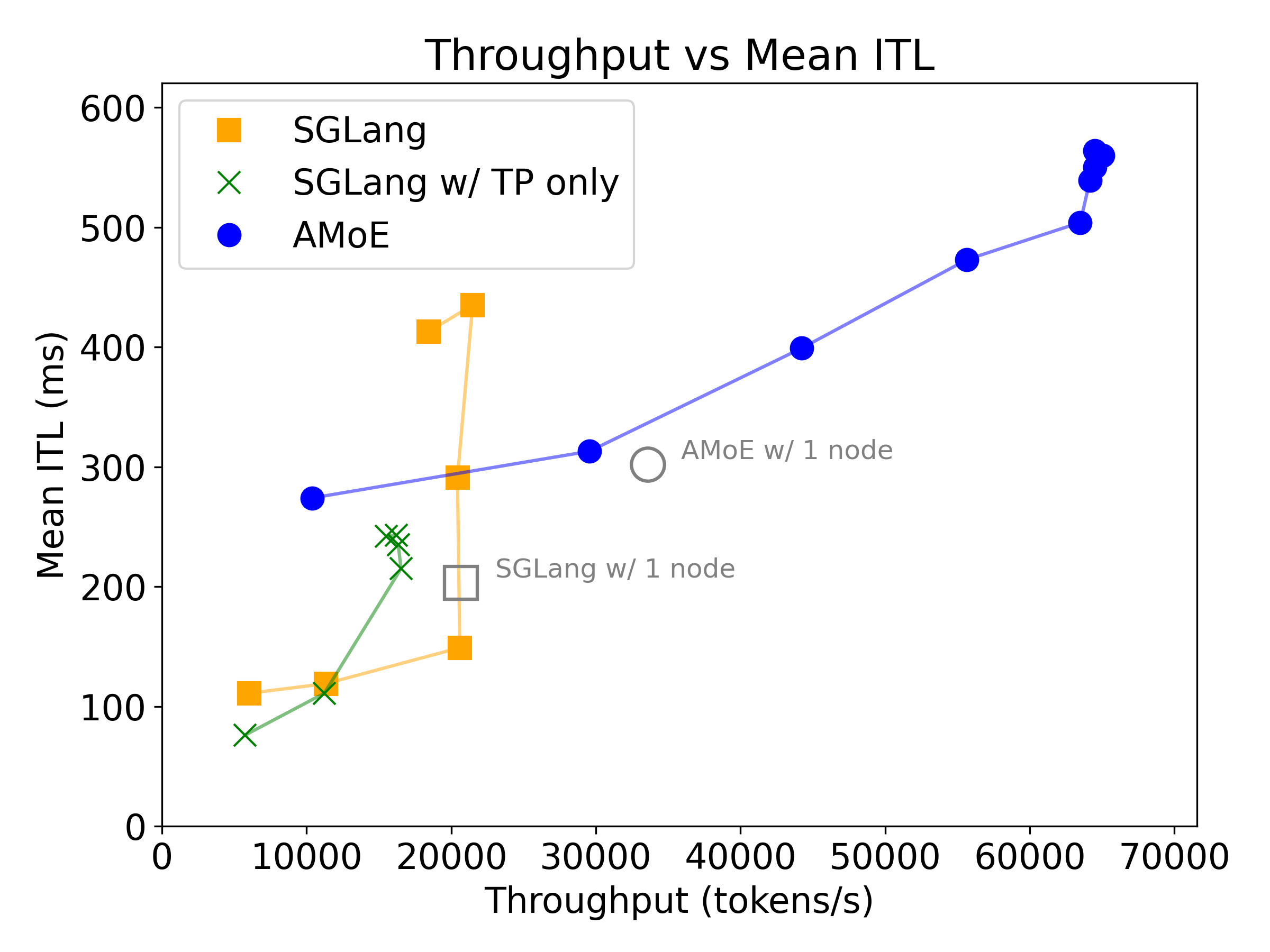}
    \caption{Performance comparison using medium workload and top-1, under a scaled setting (16 experts on 16 GPUs). The two grey-hollow markers represent the best performance under the one-node setting using AWS.}
    \label{fig:scaled-model}.
\end{figure}

We next examine how asynchronous expert parallelism performs when scaling to larger model sizes and higher parallelism, especially to a multi-node setting with datacenter networking. This benchmark is to help on predicting performance for production-scale models which has tens or hundreds of experts on many expert-parallel GPUs.

Figure~\ref{fig:scaled-model} plots the token throughput on the x-axis against the average inter-token latency (ITL) on the y-axis for a scaled-up configuration, where both the number of experts and the number of GPUs are increased (16 experts and 16 GPUs across two nodes). We use medium workload and Top-1 setting following Figure \ref{fig:mqa-top1-medium}. Compared to the baseline 8-expert setup, this expanded deployment showcases notable scalability characteristics.

When compared to SGLang with EP, \name can achieve 3x throughput improvement with a comparable ITL. The throughput gap is larger than on the single node setting with 8 experts and 8 GPUs, suggesting better scalability of \name. With more experts, systems are more susceptible to load skew, thus SGLang's throughput didn't scale even with twice number of GPUs. On the other hand, \name's throughput continues to increase, eventually achieving 1.92x improvement. 

At low input rates, \name exhibits higher ITL compared to SGLang, primarily due to the overhead introduced by layer-wise scheduling. However, as the input rate increases, the latency of \name remains relatively stable, demonstrating its ability to handle a larger volume of requests. We continue to raise the input rate until the throughput reaches saturation. Ultimately, the ITL stabilizes, indicating that the system’s performance is bounded by the available KV cache capacity.

\subsection{Efficacy of defragging scheduler}
\label{s:eval-scheduler}

\begin{figure}[t]
    \centering
    \includegraphics[width=0.9\linewidth]{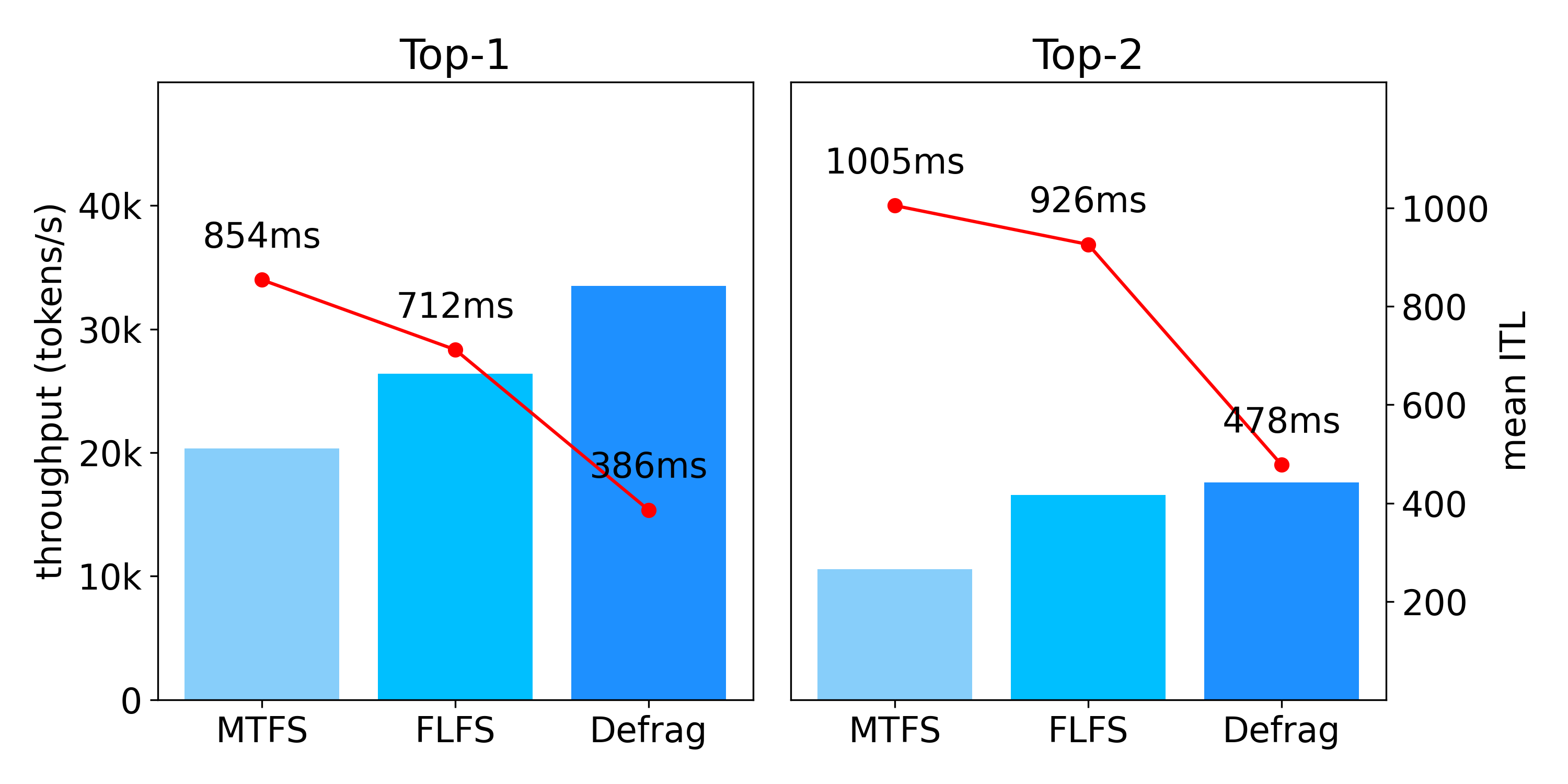}
    \caption{Compare different schedule policy at 80\% throughput. Using MQA, medium length. across experts. Top-1: input rate 250. Top-2h: input rate 150.
    }
    \label{fig:scheduler-ablation-load80}
\end{figure}

\begin{figure}[t]
    \centering
    \includegraphics[width=0.9\linewidth]{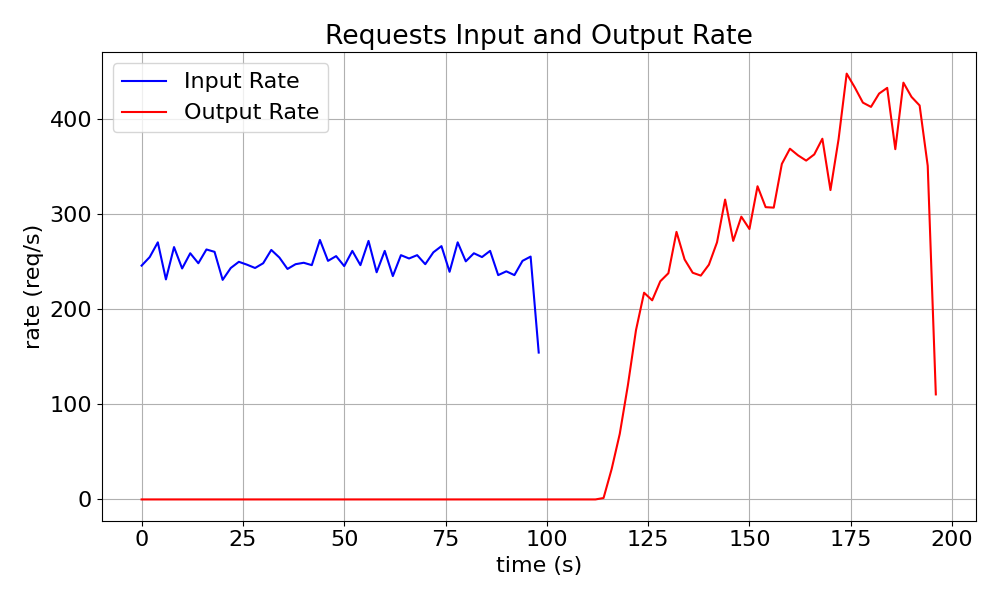}
    \caption{Request input and output rate for FLFS at input rate 250 under MQA, top-1.}
    \label{fig:flfs-livelock}
\end{figure}

We now assess the impact of our proposed defragging scheduler on system throughput and latency under different routing strategies, focusing on how the scheduler balances batch fragmentation and forward progress.

Figure~\ref{fig:scheduler-ablation-load80} compares these scheduling policies when the system operates at approximately 80\% of its maximum achievable throughput with defragging scheduler for two routing modes: Top-1 (left) and Top-2 (right). Defragging scheduler effectively minimizes batch fragmentation and shows both lower ITL and higher throughput than most-token-first-serve (MTFS) and first-layer-first-serve (FLFS) strategies.
 
Under FLFS, the scheduler always prioritizes lower-numbered blocks (i.e., those earlier in the decoding sequence) whenever there are tokens waiting in those queues. This approach effectively minimizes batch fragmentation but can severely interrupt higher-block progress, especially if new requests arrive continuously. 

Figure~\ref{fig:flfs-livelock} illustrates how FLFS can struggle when new requests keep arriving for Top-1 routing. We plot both the request input rate and the system’s request completion rate over time. Notice that once FLFS focuses on an early block for a batch of tokens, new arrivals preempt later blocks’ tokens, creating persistent waiting at higher block layers. As a result, the output rate falls behind even though the input rate remains steady. By contrast, our defragging scheduler more evenly coordinates resource usage across blocks and prevents long stalls at higher block layers, increasing the output rate closer to the input rate. 






















\subsection{Execution breakdown}
\label{section:execution-breakdown}

We conduct an in-depth investigation of the overheads introduced by layer-wise scheduling and execution. In \name, each execution step comprises five primary stages: 
\begin{itemize} 
    \item \textbf{Schedule.} The scheduler inspects the current queues and selects one layer for execution. It drains the queue and merges all tokens into one batch. 
    \item \textbf{Page Table.} The table manager allocates new key-value slots for incoming tokens and retrieves existing KV cache pages for each token. This stage is absent in expert layers. 
    \item \textbf{Pre-processing.} The executor prepares the necessary data for execution, including transferring data from CPU to GPU memory. 
    \item \textbf{Execution.} The executor launches the appropriate kernels to process a batch of tokens. 
    \item \textbf{Post-processing.} Selected outputs from execution are transferred from GPU to CPU memory to provide routing information for the dispatcher. Tokens are permuted according to either expert indices or attention data parallel indices to facilitate batch transmission. 
\end{itemize}

\begin{figure}
    \centering
    \includegraphics[width=\linewidth]{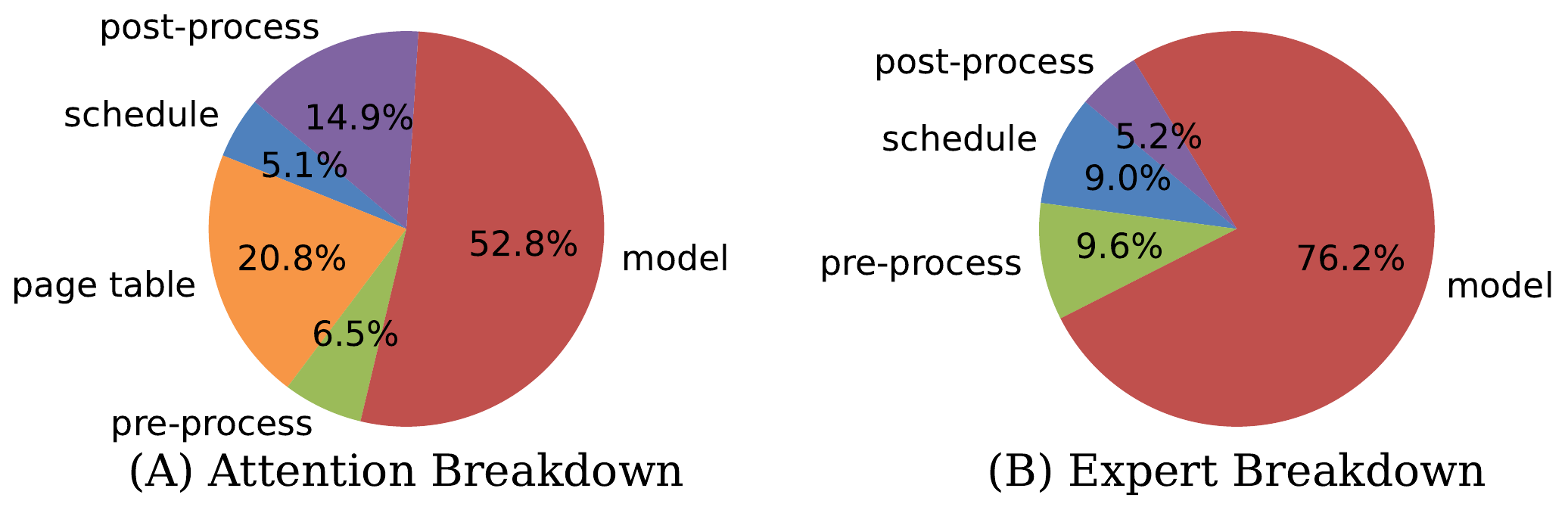}
    \caption{The attention execution takes 2.7 ms in total while the expert execution takes 0.8 ms.}
    \label{fig:execution-breakdown-merged}
\end{figure}

We sample one attention execution step and one expert execution step from a benchmark with medium workloads in Figure~\ref{fig:mqa-top1-medium} and analyze the cost of each stage. The attention step takes 2.7 milliseconds, while the expert step takes 0.8 milliseconds. Figure \ref{fig:execution-breakdown-merged} presents a detailed breakdown of an execution step for both attention and expert layers. In the attention worker, page table management incurs significant overhead, as each token must traverse the page table to retrieve its associated pages. Post-processing is also costly, as expert indices and corresponding weights must be transferred from GPU to CPU for dispatching following attention execution. However, as the length of each token’s KV cache increases, the cost of page table operations and post-processing diminishes relative to attention execution, which eventually dominates the execution step.

In contrast, during expert layer execution, the majority of time is spent on kernel computation. This is because expert kernels operate independently of metadata, and no results need to be transferred back to the CPU for subsequent dispatching, resulting in minimal post-processing overhead.

Across both attention and expert layers, the scheduling stage accounts for only a small fraction of the total time. We attribute this efficiency to our highly optimized C++ and CUDA implementation, which manages batches and token hidden states with minimal overhead.


\section{Related Works}
\subsection{MoE Serving Systems}
Several systems have been proposed to optimize the serving performance of MoE models to realize their reduced per-token computational cost~\cite{deepseekv3,deepspeedmoe,lina,exflow}. DeepSpeed-MoE~\cite{deepspeedmoe} extends expert parallelism by introducing fine-grained expert sharding, where each expert can be further divided into smaller slices, and proposes hierarchical all-to-all communication to support this finer granularity. Although this approach improves scalability with a large number of accelerators, it remains vulnerable to load imbalance caused by expert skew.

To mitigate this issue, DeepSeek-MoE~\cite{deepseekv3}, Lina~\cite{lina}, and ExFlos~\cite{exflow} leverage profiling of token routing patterns. DeepSeek-MoE and Lina duplicates frequently accessed (hot) experts based on the profiled token distribution, while ExFlos rearranges expert placement to maximize inter-layer token locality and reduce all-to-all communication volume. Despite these optimizations, existing systems continue to suffer from strict synchronization barriers imposed by all-to-all communications and the under-utilization of accelerators due to suboptimal batching in cold experts. In contrast, \name breaks this synchronization barrier and achieves higher GPU utilization through asynchronous expert parallelism.

Another line of work focuses on enabling efficient MoE serving under memory-constrained environments~\cite{moeoffloading,moecaching,pregatedmoe,readme}. To address the expanded memory footprint of MoE models, MoE-Offloading~\cite{moeoffloading} places experts in CPU memory and dynamically fetches the necessary experts to the GPU for execution. MoE-Infinity~\cite{moecaching} exploits the temporal locality of accessed experts by caching a frequently used subset on the GPU. Further advancements, such as Pregated-MoE~\cite{pregatedmoe} and ReadME~\cite{readme}, introduce new router designs that provide expert routing information for future layers ahead of time, thereby enhancing the efficiency of fetching and caching. While these works primarily target small-scale deployments where the entire model cannot fit in GPU memory, their co-design principles present interesting future directions for improving the memory efficiency of \name.

\subsection{LLM Serving Systems}
Beyond MoE-specific solutions, general LLM serving frameworks such as Orca~\cite{orca}, vLLM~\cite{vllm}, Sarathi-Serve~\cite{sarathiserve}, and Llumnix~\cite{llumnix} focus on efficient request batching and scheduling to reduce serving latency. These approaches achieve high throughput via aggressive batch formation, but they rely on strict batching and incur overheads due to bulk collective communication, which can undermine latency benefits. Disaggregated serving designs like DistServe~\cite{distserve} and SplitWise~\cite{splitwise} instead decouple the prefill and decoding phases to handle their differing performance characteristics, concentrating mainly on optimizing decoding throughput and latency. However, while disaggregation mitigates prefill–decoding interference and boosts decode-phase efficiency, these systems do not consider the unique challenges of MoE models, such as expert load imbalance and inefficient token communications.
\section{Conclusion}
In conclusion, our proposed Asynchronous Expert Parallelism (AEP) effectively addresses the GPU underutilization and synchronization bottlenecks that commonly arise in expert-parallel MoE serving. By introducing $\mu$-queuing and a defragging scheduler, our system \name re-batches tokens adaptively, reducing idle time and improving throughput. Evaluations indicate that \name can achieve up to 3$\times$ higher throughput than state-of-the-art baselines, with minimal latency overhead, and scales much better to multi-node configurations. Taken together, these results affirm that AEP is a promising approach to efficiently handle large-scale MoE deployments.

\clearpage
\small
\bibliographystyle{plain}
\bibliography{ref}

\clearpage
\appendix

\end{document}